\documentclass[5p,longtitle]{elsarticle}

\usepackage{graphicx}
\usepackage{amsmath,amssymb}
\usepackage{lineno}
\usepackage[dvipdfmx]{hyperref}
\usepackage[flushleft]{threeparttable}
\modulolinenumbers[5]

\journal{Nuclear Instruments and Methods in Physics Research Section A}

\begin{document}

\begin{frontmatter}

\title{Modeling of proton-induced radioactivation background in hard X-ray telescopes: Geant4-based simulation and its demonstration by \textit{Hitomi}'s measurement in a low Earth orbit}

\author[affil_kipac,affil_riken_nishina]{Hirokazu~Odaka\corref{mycorrespondingauthor}}

\address[affil_kipac]{Kavli Institute for Particle Astrophysics and Cosmology, Stanford University, 452 Lomita Mall, Stanford, CA 94305, USA}
\address[affil_riken_nishina]{Nishina Center for Accelerator-Based Science, RIKEN, 2-1 Hirosawa, Wako, Saitama 351-0198, Japan}
\cortext[mycorrespondingauthor]{Corresponding author}
\ead{hirokazu.odaka@riken.jp}

\author[affil_slac]{Makoto~Asai}
\author[affil_rikadai]{Kouichi~Hagino}
\author[affil_slac]{Tatsumi~Koi}
\author[affil_kipac,affil_slac]{Greg~Madejski}
\author[affil_hiroshima_astro]{Tsunefumi~Mizuno} 
\author[affil_hiroshima]{Masanori~Ohno}
\author[affil_rikkyo]{Shinya~Saito}
\author[affil_ut,affil_isas]{Tamotsu~Sato}
\author[affil_slac]{Dennis~H.~Wright}

\author[affil_kyoto_astro,affil_kyoto_hakubi]{Teruaki~Enoto}
\author[affil_hiroshima]{Yasushi~Fukazawa}
\author[affil_isas,affil_nagoya]{Katsuhiro~Hayashi}
\author[affil_waseda]{Jun~Kataoka}
\author[affil_hiroshima]{Junichiro~Katsuta} 
\author[affil_isas]{Madoka~Kawaharada}
\author[affil_kyoto_physics]{Shogo~B.~Kobayashi} 
\author[affil_isas]{Motohide~Kokubun} 
\author[affil_apc,affil_cea]{Philippe~Laurent} 
\author[affil_apc]{Francois~Lebrun} 
\author[affil_cea]{Olivier~Limousin} 
\author[affil_cea]{Daniel~Maier} 
\author[affil_riken_maxi]{Kazuo~Makishima} 
\author[affil_waseda]{Taketo~Mimura} 
\author[affil_ut]{Katsuma~Miyake} 
\author[affil_isas]{Kunishiro~Mori} 
\author[affil_ut]{Hiroaki~Murakami} 
\author[affil_yamagata]{Takeshi~Nakamori} 
\author[affil_riken_nishina]{Toshio~Nakano} 
\author[affil_ut,affil_ut_rescue]{Kazuhiro~Nakazawa} 
\author[affil_tohoku_frontier,affil_tohoku]{Hirofumi~Noda} 
\author[affil_isas]{Masayuki~Ohta} 
\author[affil_isas]{Masanobu~Ozaki}
\author[affil_isas]{Goro~Sato} 
\author[affil_isas]{Rie~Sato} 
\author[affil_nagoya_space]{Hiroyasu~Tajima} 
\author[affil_hiroshima]{Hiromitsu~Takahashi} 
\author[affil_isas]{Tadayuki~Takahashi} 
\author[affil_oist]{Shin'ichiro~Takeda} 
\author[affil_kyoto_physics]{Takaaki~Tanaka} 
\author[affil_hiroshima_astro]{Yasuyuki~Tanaka} 
\author[affil_saitama]{Yukikatsu~Terada} 
\author[affil_shizuoka]{Hideki~Uchiyama} 
\author[affil_rikkyo]{Yasunobu~Uchiyama} 
\author[affil_isas]{Shin~Watanabe} 
\author[affil_nagoya,affil_nagoya_space]{Kazutaka~Yamaoka} 
\author[affil_saitama]{Tetsuya~Yasuda}
\author[affil_tokyotech]{Yoichi~Yatsu} 
\author[affil_riken_nishina]{Takayuki~Yuasa} 
\author[affil_ssl]{Andreas Zoglauer}

\address[affil_slac]{SLAC National Accelerator Laboratory, 2575 Sand Hill Road, Menlo Park, CA 94025, USA}
\address[affil_rikadai]{Department of Physics, Tokyo University of Science, 2641 Yamazaki, Noda, Chiba, 278-8510, Japan}
\address[affil_hiroshima]{School of Science, Hiroshima University, 1-3-1 Kagamiyama, Higashi-Hiroshima 739-8526, Japan}
\address[affil_hiroshima_astro]{Hiroshima Astrophysical Science Center, Hiroshima University, Higashi-Hiroshima, Hiroshima 739-8526, Japan}
\address[affil_rikkyo]{Department of Physics, Rikkyo University, 3-34-1 Nishi-Ikebukuro, Toshima-ku, Tokyo 171-8501, Japan}
\address[affil_ut]{Department of Physics, The University of Tokyo, 7-3-1 Hongo, Bunkyo-ku, Tokyo 113-0033, Japan}
\address[affil_isas]{Japan Aerospace Exploration Agency, Institute of Space and Astronautical Science, 3-1-1 Yoshino-dai, Chuo-ku, Sagamihara, Kanagawa 252-5210, Japan}
\address[affil_kyoto_astro]{Department of Astronomy, Kyoto University, Kitashirakawa-Oiwake-cho, Sakyo-ku, Kyoto 606-8502, Japan}
\address[affil_kyoto_hakubi]{The Hakubi Center for Advanced Research, Kyoto University, Kyoto 606-8302, Japan}
\address[affil_nagoya]{Department of Physics, Nagoya University, Furo-cho, Chikusa-ku, Nagoya, Aichi 464-8602, Japan}
\address[affil_waseda]{Research Institute for Science and Engineering, Waseda University, 3-4-1 Ohkubo, Shinjuku, Tokyo, 169-8555, Japan}
\address[affil_kyoto_physics]{Department of Physics, Kyoto University, Kitashirakawa-Oiwake-Cho, Sakyo, Kyoto 606-8502, Japan}
\address[affil_apc]{Laboratoire APC, 10 rue Alice Domon et L\'eonie Duquet, 75013 Paris, France}
\address[affil_cea]{CEA Saclay, 91191 Gif-sur-Yvette, France}
\address[affil_riken_maxi]{Maxi Team, RIKEN, 2-1 Hirosawa, Wako, Saitama 351-0198, Japan}
\address[affil_yamagata]{Faculty of Science, Yamagata University, 1-4-12 Kojirakawa-machi, Yamagata, Yamagata 990-8560, Japan}
\address[affil_ut_rescue]{Research Center for the Early Universe, School of Science, The University of Tokyo, 7-3-1 Hongo, Bunkyo-ku, Tokyo 113-0033, Japan}
\address[affil_tohoku_frontier]{Frontier Research Institute for Interdisciplinary Sciences, Tohoku University,  6-3 Aramakiazaaoba, Aoba-ku, Sendai, Miyagi 980-8578, Japan}
\address[affil_tohoku]{Astronomical Institute, Tohoku University, 6-3 Aramakiazaaoba, Aoba-ku, Sendai, Miyagi 980-8578, Japan}
\address[affil_nagoya_space]{Institute for Space-Earth Environmental Research, Nagoya University, Furo-cho, Chikusa-ku, Nagoya, Aichi 464-8601, Japan}
\address[affil_oist]{Okinawa Institute of Science and Technology Graduate University, 1919-1 Tancha, Onna-son Okinawa, 904-0495, Japan}
\address[affil_saitama]{Department of Physics, Saitama University, 255 Shimo-Okubo, Sakura-ku, Saitama, 338-8570, Japan}
\address[affil_shizuoka]{Faculty of Education, Shizuoka University, 836 Ohya, Suruga-ku, Shizuoka 422-8529, Japan}
\address[affil_tokyotech]{Department of Physics, Tokyo Institute of Technology, 2-12-1 Ookayama, Meguro-ku, Tokyo 152-8550, Japan}
\address[affil_ssl]{Space Sciences Laboratory, University of California, Berkeley, CA 94720, USA}

\begin{abstract}
Hard X-ray astronomical observatories in orbit suffer from a significant amount of background due to radioactivation induced by cosmic-ray protons and/or geomagnetically trapped protons. Within the framework of a full Monte Carlo simulation, we present modeling of in-orbit instrumental background which is dominated by radioactivation. To reduce the computation time required by straightforward simulations of delayed emissions from activated isotopes, we insert a semi-analytical calculation that converts production probabilities of radioactive isotopes by interaction of the primary protons into decay rates at measurement time of all secondary isotopes. Therefore, our simulation method is separated into three steps: (1) simulation of isotope production, (2) semi-analytical conversion to decay rates, and (3) simulation of decays of the isotopes at measurement time. This method is verified by a simple setup that has a CdTe semiconductor detector, and shows a 100-fold improvement in efficiency over the straightforward simulation. To demonstrate its experimental performance, the simulation framework was tested against data measured with a CdTe sensor in the Hard X-ray Imager onboard the \textit{Hitomi} X-ray Astronomy Satellite, which was put into a low Earth orbit with an altitude of 570 km and an inclination of 31$^\circ$, and thus experienced a large amount of irradiation from geomagnetically trapped protons during its passages through the South Atlantic Anomaly. The simulation is able to treat full histories of the proton irradiation and multiple measurement windows. The simulation results agree very well with the measured data, showing that the measured background is well described by the combination of proton-induced radioactivation of the CdTe detector itself and thick $\mathrm{Bi_4Ge_3O_{12}}$ scintillator shields, leakage of cosmic X-ray background and albedo gamma-ray radiation, and emissions from naturally contaminated isotopes in the detector system.
\end{abstract}


\end{frontmatter}


\section{Introduction}


Hard X-ray telescopes for astrophysics must be in orbit because of atmospheric absorption, and therefore suffer from 
significant backgrounds induced by cosmic rays and/or geomagnetically trapped charged particles.
Direct ionization signals of the charged particles and prompt gamma-ray emissions they cause can be eliminated by 
the anti-coincidence of active shields associated with primary detectors \cite{Takahashi:2007}.
However, delayed emissions from radioactive isotopes produced by interactions with detector material of cosmic-ray protons and/or geomagnetically trapped protons remain\cite{Kokubun:2007, Wik:2014}.
These kinds of background may arise from the inside of the detectors themselves and therefore are extremely difficult to 
eliminate as noise since neither anti-coincidence nor collimators are effective for rejecting it.  Thus, evaluation 
of the radioactivation background via modeling must be an important performance factor for hard X-ray observations.


Monte Carlo simulation has been an effective means of evaluating the background in a phase of mission planning since 
the instrumental design and the selection of orbit must depend on an estimate of the background 
\cite{Weidenspointner:2005, Zoglauer:2006, Mizuno:2010, Ozaki:2012}.
It provides crucial information for optimizing the selection of detector material, arrangement of the detectors, 
shields and their supporting structure, and data acquisition strategies.
Simulation is also necessary for estimating the background before the launch of the satellite, allowing us to develop a 
specific science program (astronomical observation planning).


In the data analysis phase phenomenological methods, rather than the full simulation, have typically been employed for 
the purpose of subtracting the background from the obtained data primarily because empirical modeling based on measured 
data has been considered sufficiently accurate and even more efficient.  In addition, due to the complexity of hadronic 
processes, simulations have not necessarily achieved sufficient model accuracy.
Nonetheless, a full simulation is a more promising approach because of its comprehensiveness and flexibility.
Empirical methods are normally limited by various 
conditions such as energy bands, observation time windows, or fields of view.  They are further limited by data quality 
and availability.  Demands for comprehensive simulations will increase as instruments achieve higher resolutions and more 
complex structures both in hardware and software (e.g, multiple layers, a number of readout channels, and a variety of 
data acquisition modes).


The full simulation of radioactivation is a significantly complicated process.
In addition to the difficulties associated with normal Monte Carlo simulations, such as inaccuracies of mass models, 
physical process implementation, and input radiation environments in orbit, it requires treatment of the delayed 
nature of radioactivation.  Thus, the simulation must consider the entire history of the particle irradiation and 
integrate events over a time window of interest.  MGGPOD \cite{Weidenspointner:2005} introduced an efficient scheme to 
solve this problem by separating the calculation into the production of radioactive isotopes and the decays of these 
isotopes.  To connect these two phases, it is necessary to convert the production information into the decay rates of 
the isotopes, which can be done by analytical or numerical methods; it does not require time-consuming Monte Carlo 
calculations.


We have developed a new general-purpose simulation framework to evaluate radioactivation in orbit by adopting the 
above scheme.  Our framework utilizes the \textsc{Geant4} toolkit library\cite{Agostinelli:2003, Allison:2006, Allison:2016},
which is a widely used Monte Carlo simulation package, allowing full compatibility with other types of important 
simulations including photon signals and other backgrounds such as cosmic X-ray background, Earth's albedo gamma rays and 
neutrons, prompt emissions due to cosmic rays, and so on.
We used \textsc{Geant4}, version 10.04.b01, in order to apply the latest hadronic physics models and associated 
databases including nuclear tables.
Our code treats an arbitrary irradiation time profile in order to account for the highly variable radiation environment along the orbit.


For software verification and performance evaluation, we compared simulation results to in-orbit data obtained with the Hard X-ray Imager (HXI) \cite{Nakazawa:2016, Nakazawa:2017} onboard the \textit{Hitomi} X-ray astronomy satellite \cite{Takahashi:2016, Takahashi:2017}, which was put into a low Earth orbit (LEO).
A spectrum measured by the HXI is highly suitable for this test in the energy range of 10--160 keV; the detector materials, one of which is cadmium telluride (CdTe) in its main focal plane imager, had been exposed to highly variable, high flux geomagnetically trapped protons during their passages through Earth's radiation belt, generating instrumental background dominated by proton-induced radioactivation.
The Soft Gamma-ray Detector (SGD) \cite{Watanabe:2016,Tajima:2017} aboard the same satellite also provided us with useful data at higher energies up to 600 keV, though we focus on the HXI background in this paper because of 
less complicated structure and data reduction methods of the HXI.
Data analysis of the SGD background using our simulation framework will be presented separately in future.


Neutrons generated by interactions of cosmic rays with Earth's atmosphere can also cause background that is difficult 
to reject by the anti-coincidence technique.  A hard X-ray detector is sensitive to these atmospheric neutrons mainly via 
elastic scattering inside the detector itself, prompt emissions after inelastic interactions, and delayed emissions due 
to radioactivation.  Our simulation framework is also applicable to the radioactivation induced by neutrons.
In the case of the HXI, however, the background of the CdTe detector is dominated by radioactivation due to trapped 
protons, and the contribution from atmospheric neutrons is negligible.  It should be noted that the silicon (Si) 
detectors of the HXI and SGD are influenced by background produced via neutron elastic scattering by Si nuclei 
\cite{Mizuno:2010, Tajima:2017, Hagino:2017}.

In this paper, we describe a method for accurately modeling the instrumental background of a hard X-ray telescope using
a full Monte Carlo simulation.  Section \ref{sec:method} describes the framework and the method of the simulation.
In Section \ref{sec:verification}, a simple case study using a CdTe detector is presented for verification of the method and 
the software code.
Section \ref{sec:measured_data} describes the HXI and its measured data used for testing the simulation.
In Section \ref{sec:results} we present concrete methods and conditions of the simulations of the HXI background, their 
results, and their comparison with data.
In Section \ref{sec:discussion} we confront the results of our simulations with experimental data, and discuss possible improvements.
This is followed by our conclusions in Section \ref{sec:conclusions}.

\section{Methodology}
\label{sec:method}

\begin{figure*}[t]
\begin{center}
\includegraphics[width=15cm]{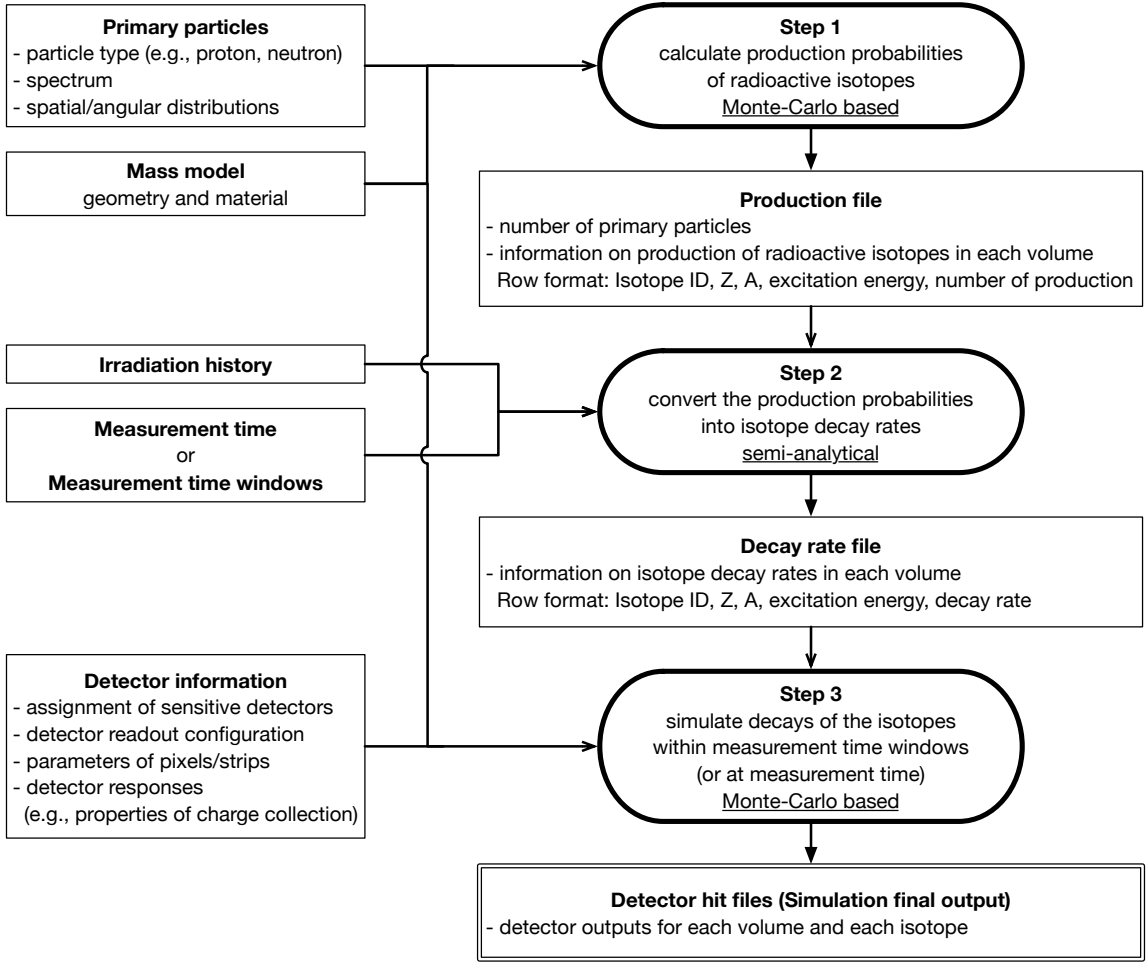}
\caption{Overview of the calculation.}
\label{fig:overview}
\end{center}
\end{figure*}

As already mentioned, the delayed nature of radioactivation is an important issue for the simulation.
It would be in principle possible to perform a straightforward simulation that starts with primary (incident) particles, such as 
cosmic-ray protons, and obtain final beta-ray and gamma-ray emissions.  However, the resultant emissions would be widely 
distributed in the temporal domain, requiring a large number of Monte Carlo trials to collect statistically sufficient event samples in a specific time window.
Instead of using Monte Carlo simulations for the entire process, we can reduce computation time by exploiting analytical 
or numerical calculations of the decay chains which link the isotope production and decays within the observation time.
This procedure was originally introduced by MGGPOD \cite{Weidenspointner:2005} and was also adopted by MEGAlib 
\cite{Zoglauer:2006}.
Employing this scheme, we define three steps to calculate the resultant background events due to radioactivation: 
\begin{enumerate}
\item simulate the production of radioactive isotopes by the interaction of primary particles with assumed input spectra, 
\item calculate decay rates at the measurement time of these generated isotopes and their descendants, and
\item simulate their decays and the interactions of the associated secondary particles.
\end{enumerate}
Figure~\ref{fig:overview} illustrates the structure of these three steps in the simulation framework.

Our framework is based on the \textsc{Geant4} toolkit library \cite{Agostinelli:2003, Allison:2006, Allison:2016}, using 
version 10.04.b01 for this work.  \textsc{Geant4} performs the Monte Carlo simulations in steps 1 and 3.
Step 2, while not a Monte Carlo simulation but a semi-analytical calculation, does use the \textsc{Geant4} library and its 
attached database to extract decay chains of radioactive isotopes, yielding fully consistent results with the 
straightforward Monte Carlo simulation.

The purpose of Step 1 is to obtain production probabilities of radioactive isotopes by a primary particle irradiation.
This step is a normal Monte Carlo simulation in which we focus only on secondary radioactive nuclei generated immediately 
after inelastic interactions of the primary protons with nuclei in the target material.  In each simulated \textit{event}, 
which corresponds to a single primary particle generation, the code records information on the secondary nucleus (if it exists) by detecting its decay process.
The information stored here includes the atomic number $Z$, the atomic mass number $A$ and 
the excitation energy $E$ of this nucleus, the interaction position and the physical volume.
After the generation of a secondary nucleus, the code stops tracking the generated isotope and all its secondaries in order to avoid double-counting its descendants.
Throughout the entire simulation run, we sum up all generated isotopes in each physical volume of 
the mass model, producing a summary file in ascii format (\textit{production file} shown in Figure~\ref{fig:overview}).

The special treatments of Step 1 are implemented in the \textit{stepping action} phase, in which \textsc{Geant4} allows application developers to supply user-defined actions to be invoked at every step of particle tracking.
In the code, the generation of an isotope is detected by checking whether a specific physical process responsible for the radioactive decay, which is \texttt{G4RadioactiveDecay}, is invoked.
After storing this isotope, the processes which track it and all its secondaries are killed.
We need to collect isotopes that have sufficiently long lifetimes to produce the delayed emissions, and therefore the code should allow subsequent radioactive decays of relatively short-lived isotopes.
The algorithm for the \textit{stepping action} is represented by the following pseudo code: 
\begin{verbatim}
if process_name is "RadioactiveDecay"
  if lifetime >= lifetime_limit
    record the isotope information (Z, A, E...)
    kill this track and all its secondaries
  else
    do nothing (allow following tracking)
  end
end
\end{verbatim}
The lifetime limit is a changeable parameter to determine if the isotope should be collected, and is normally set to
1 millisecond.

Step 2 performs a semi-analytical calculation of radiative decay, converting the isotope production probabilities into 
decay rates of related isotopes at a time of interest.
Consider that a radioactive isotope $X_0$ has a decay chain:
\begin{equation}\label{eq:chain}
X_0 \xrightarrow[f_1]{\lambda_1} X_1 \xrightarrow[f_2]{\lambda_2} X_2 \xrightarrow[f_3]{\lambda_3}\cdots  \xrightarrow[]{}X_{n-1} \xrightarrow[f_{n}]{\lambda_{n}} X_{n},
\end{equation}
where $X_i$ is the $i$-th isotope in the chain, and $\lambda_i$ and $f_i$ denote the decay constant (inverse of the lifetime) and the branching ratio, respectively.
The equation\footnote{The indexing of the decay constant is based on the index of a link (decay) between a parent nucleus and its daughter as indicated in Equation (\ref{eq:chain}), though it is usually considered to be a property of the parent.} describing the decay chain (the Bateman equation \cite{Bateman:1910, Harr:2007}) is given by
\begin{gather}
\frac{dn_0}{dt} = -\lambda_1 n_0, \\
\frac{dn_i}{dt} = \lambda_{i} n_{i-1} - \lambda_{i+1} n_i \quad (i\ge 1),
\end{gather}
where $n_i$ is the number of the $i$-th isotope $X_i$.
The analytic solution is given by
\begin{gather}
n_0(t) = n_0(0) \exp(-\lambda_1 t)\;(t\ge 0),\;0\;(t<0), \label{eq:solution1} \\
n_i(t) = n_0(0) \sum_{j=1}^{i+1} D_{ij} \exp(-\lambda_j t)\;(t\ge 0),\;0\;(t<0), \label{eq:solution2}
\end{gather}
where $n_i(t)$ is the number of isotopes $X_i$ as a function of time $t$, assuming that the time profile of the proton irradiation is described by the delta function $\delta(t)$, and $n_0(0)$ is the number of isotopes $X_0$ generated by the instantaneous irradiation---in other words, the production probability of the isotope generation by a single irradiation; the coefficients $D_{ij}$ are given by
\begin{gather}
D_{ij} = \prod_{p=1}^{i+1} A_p B_{jp}, \\
A_p = \lambda_p \;(p=1\ldots i),\; 1\;(p=i+1), \\
B_{jp} = \frac{1}{\lambda_p-\lambda_j} \;(p\neq j),\; 1\;(p=j).
\end{gather}

For a given time profile $g(\tau)$ of irradiation, the number of isotopes $X_i$ can be written as 
\begin{equation}\label{eq:convolution}
N_i(t) = \int_{\tau=-\infty}^{\tau=+\infty} g(\tau) n_i(t-\tau)d\tau.
\end{equation}
Any time profile $g(\tau)$ of irradiation can be written as the sum of a number of constant profiles:
\begin{gather}
g(\tau) = \sum_{m} g_{m}(\tau) \label{eq:sum_time_profile},\\
g_{m}(\tau) = g_{m}\;(\tau_{m1} < \tau < \tau_{m2}),\;0\;(\mathrm{otherwise}),
\end{gather}
where each time section indexed by $m$ ranges between $\tau_{m1}$ and $\tau_{m2}$.
So Equation~(\ref{eq:convolution}) can be written as
\begin{equation}
N_i(t) = \sum_m g_m \int_{t=\tau_{m1}}^{t=\tau_{m2}} N_i(t-\tau)d\tau,
\end{equation}
in which the integral is performed analytically.
If the observation is performed in the time window between $t_1$ and $t_2$, the time-averaged population of the isotope 
is given by
\begin{equation}
\overline{N_i} = \frac{1}{t_2-t_1}\int_{t_1}^{t_2} N_i(t) dt.
\end{equation}

Taking account of the branching ratios at all the decay steps, the solution $N_i(t)$ should be weighted by a factor of
\begin{equation}
w = \prod_i f_i = f_1 f_2 \cdots f_{n},
\end{equation}
and should also be summed over all decay chains which include this isotope, as
\begin{equation}
N_i(t) = \sum_{k\in\mathrm{chains}} w^{(k)} N_i^{(k)}(t),
\end{equation}
where $k$ is a label of a decay chain.
The decay rate is simply given by multiplying it by the decay constant of the isotope.

Now we have obtained the decay rates of all isotopes in all volumes of interest at a specific time or in a specific time 
window.  Step 3 then conducts full Monte Carlo simulations of the decays of these isotopes.
Each simulation \textit{event} starts with a specific radioactive isotope as a primary particle put inside a specific volume.
The position of the primary is uniformly sampled in the physical volume as this treatment gives a sufficiently good 
approximation of the space environment.

These steps are packaged into \textit{ComptonSoft}\footnote{https://github.com/odakahirokazu/ComptonSoft}, which is an 
integrated simulation and analysis software suite for semiconductor radiation detectors including Compton cameras 
\cite{Odaka:2010}.  This package together with a mass model of the HXI has been used for the detector design and 
performance evaluation of the HXI.  While \textit{ComptonSoft} is already matched to multi-purpose applications, we will 
provide standalone software programs for each step since the steps are only loosely coupled to one another.

\section{Code Verification}
\label{sec:verification}

\begin{figure*}[t]
\begin{center}
\includegraphics[width=18cm]{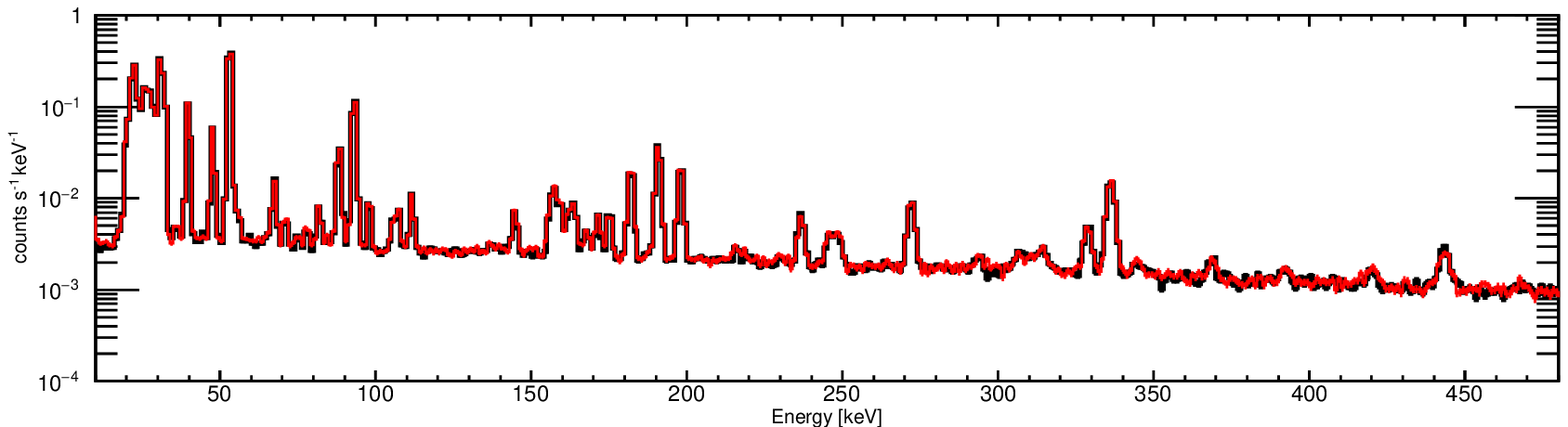}
\includegraphics[width=18cm]{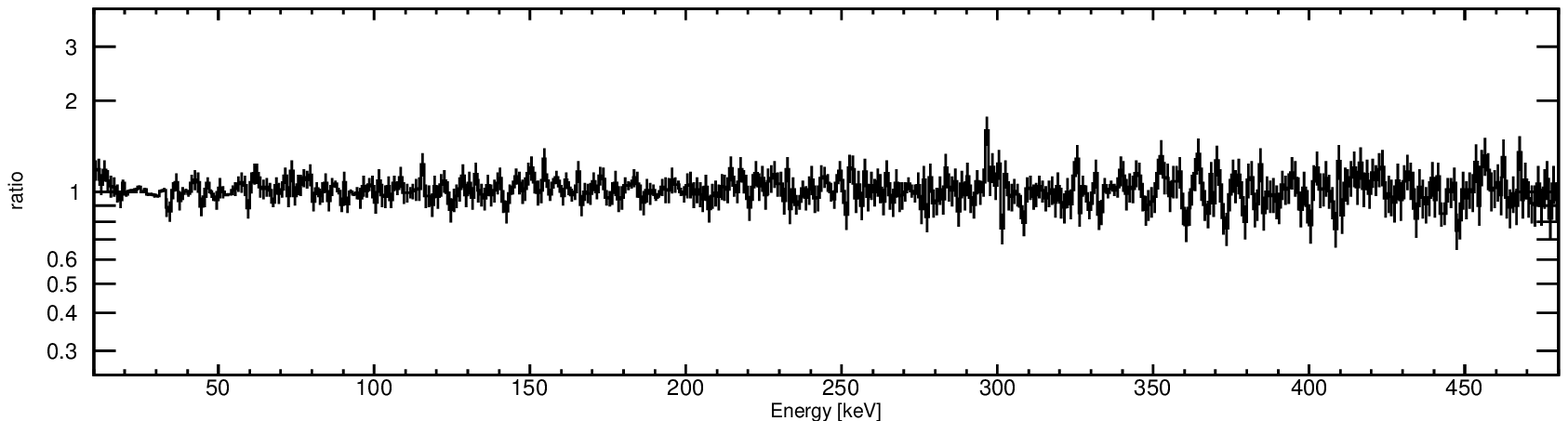}
\caption{Top: comparison of CdTe radioactivation spectrum between the efficient, three-step approach (red) and the
 straightforward approach (black). Bottom: ratio of the spectra from the three-step approach to that of the straightforward approach.
 Note that the error bars in both panels do not include statistical errors due to Step 1 of the three-step approach (see text).}
\label{fig:verification}
\end{center}
\end{figure*}

\begin{table}[t]
\caption{Computation time of the simulations}
\label{table:computation_time}
\begin{center}
\begin{tabular}{crl}
\hline\hline
Simulation & CPU time & Note \\
\hline
Straightforward & 2410000 s & $10^{10}$ events\\
Step 1 & 13900 s & $10^8$ events \\
Step 2 & 0.5 s & semi-analytical \\
Step 3 & 478 s & 560059 events \\
\hline
\end{tabular}
\end{center}
\end{table}%

In this section, we present a simple case of proton-induced radioactivation in order to verify our simulation method.
We compare two methods: (1) a straightforward Monte Carlo simulation that tracks all processes ranging from the primary 
protons to decays of all isotopes associated with the primary interactions, and (2) the three-step approach introduced 
in the previous section.  The first method is a normal \textsc{Geant4} simulation run. 
Both should agree within statistical errors if we use completely identical physical process implementations and nuclear databases.

For the comparison we used a simple simulation setup in order to reduce the CPU time required by the straightforward approach. 
We considered a pixel detector of CdTe semiconductor since this material was used for the experimental test in this work (See \S\ref{sec:measured_data}).
The detector has a sensitive area of $25.6\times 25.6\;\mathrm{mm^2}$ which is segmented into $8\times 8$ pixels.
Gaussian noise was added to each hit signal with a standard deviation (1$\sigma$) of $\Delta E = \sqrt{(0.4\;\mathrm{keV})^2 + (0.003E)^2}$.
The detector was irradiated by a monochromatic 150 MeV proton beam with a radius of 20 mm, which covered the entire detector wafer.
The irradiation was assumed to be instant (like a delta function) with a total number of $10^{10}$ 
primary protons, and then we measured delayed signals within a time window from $1.0\times 10^5$ s to $2.0\times 10^5$ after the irradiation.

We used the \texttt{G4BinaryCascade} hadronic physics model, which is part of the \texttt{QGSP\_BIC\_HP} physics list.
The \texttt{G4Radio\-active\-Decay} process was then added to this physics list, allowing decays of generated secondary isotopes.
This physics setup is common to all the simulations in this section.  The nuclear databases 
\texttt{Radioactive\-Decay\-5.1.1} and \texttt{Photon\-Evaporation\-5.0.2} were used here and in Step 2.

Straightforward simulations were conducted with $10^{10}$ primary protons, matching the assumed number of the irradiation.
From the output event list, we selected events detected within the measurement time window, and extracted a spectrum in an 
energy range of 10--480 keV.  A spectrum under the same conditions was obtained by the three step approach.
In Step 1, we simulated $10^8$ events to accumulate the isotopes generated by the primary interactions.
This number of the simulated events was chosen in order to have sufficient statistics of the generated nuclei in the Step 1 output.
Although the number of primaries was 100 times smaller than that of the straightforward run, this should be sufficient 
since this method does not waste events at the initial irradiation to obtain the final result; on the other hand, most of the events in the straightforward approach result in making signals outside of the time window of interest.
In Step 2 the average decay rates were calculated within the time window, assuming the instantaneous (very short) 
irradiation of $10^{10}$ protons.  Step 3 then simulated all these decays according to the output of Step 2.

Figure~\ref{fig:verification} shows the spectra obtained by the two different methods, together with their ratio.
For the three-step method, error bars in these plots indicate only statistical errors in the final step (Step 3) of the Monte Carlo simulations, and therefore the spectrum has hidden statistical errors due to Step 1, which is discussed in the next paragraph.
Both methods agree very well on the line energies, line intensities, and continuum levels.
This agreement guarantees that we are able to replace the normal, time-consuming simulation of the delayed radioactivation emissions with the much more efficient three-step method.
The statistical fluctuations get larger at high energies since counts per spectral bin decrease with energy.
An example of the computation time for these simulations is given in Table~\ref{table:computation_time}, showing the greatly improved efficiency of the three-step 
simulation.
These CPU times were measured on an Apple iMac (27-inch, late 2015) with Intel Core i7 (skylake) CPU 4~GHz operating in single thread mode.

Finally, we discuss the reduction of events to be simulated in the three-step approach.
The number of events to be simulated is determined by how small one wants to have statistical errors due to the Monte Carlo simulations.
Compared with the statistical errors of the straightforward method, those of the three-step simulations are more complicated as they come from the two Monte Carlo stages, namely, Step 1 and Step 3.
Although we can estimate the statistical errors of the isotope production rates in the Step 1 simulation, it is not trivial to estimate the number of events that is required for enough statistics, which largely depends on the decay properties of the isotopes (particularly the decay constants) and the measurement time windows.
A practical method to evaluate the statistical errors due to Step 1 is to compare multiple (at least two) simulation sets that have different initial random seeds, and we suggest that we should have a sufficient number of events so that the discrepancies between the different simulation sets are comparable to or smaller than statistical errors due to Step 3.

\section{Measured Data}
\label{sec:measured_data}

\begin{figure}[t]
\begin{center}
\includegraphics[width=5.5cm]{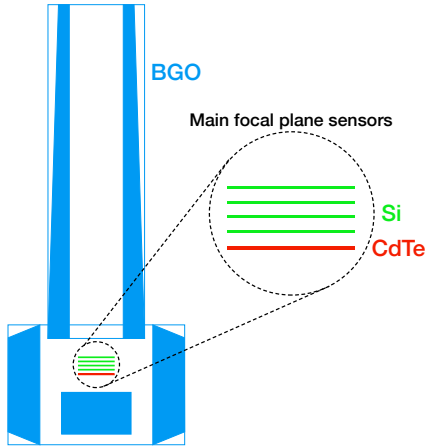}
\caption{Schematic drawing of the HXI. Only active detectors including anti-coincidence BGO shields are shown. The main focal plane imager consists of four layers of Si and one layer of CdTe double-sided strip detectors.}
\label{fig:hxi_drawing}
\end{center}
\end{figure}

\begin{figure}[t]
\begin{center}
\includegraphics[width=8cm]{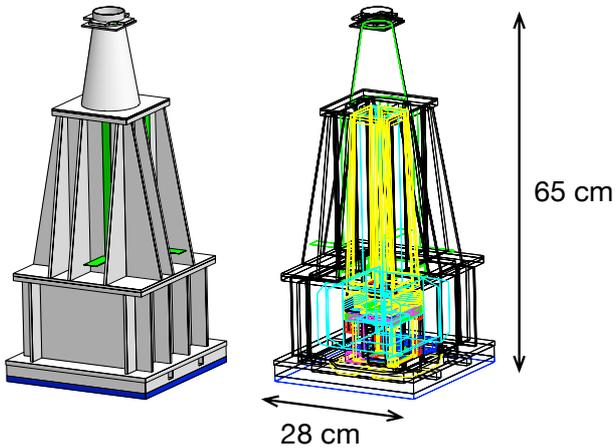}
\caption{Mass model of the HXI system. The wireframe representation is shown in the right panel.}
\label{fig:mass_model}
\end{center}
\end{figure}

\begin{figure}[t]
\begin{center}
\includegraphics[width=9cm]{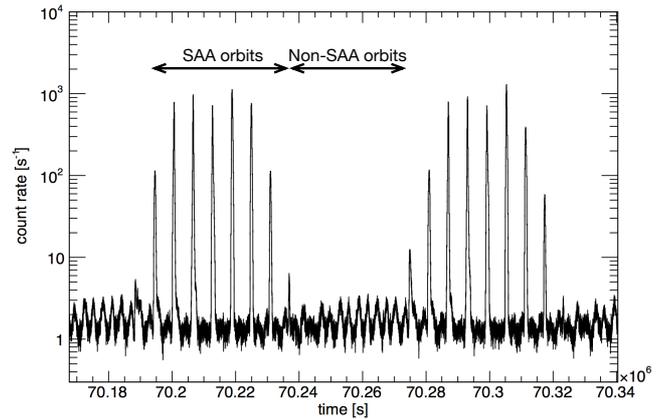}
\caption{Count rate of the particle monitor as a function of time within 48 hours.  Time is indicated as the 
         \textit{Hitomi} satellite standard time, which measures from 2014-01-01:00:00:00 UTC.  An example of the 
         classification of the SAA/non-SAA orbits is also indicated.}
\label{fig:particle_monitor}
\end{center}
\end{figure}

\begin{figure}[t]
\begin{center}
\includegraphics[width=9cm]{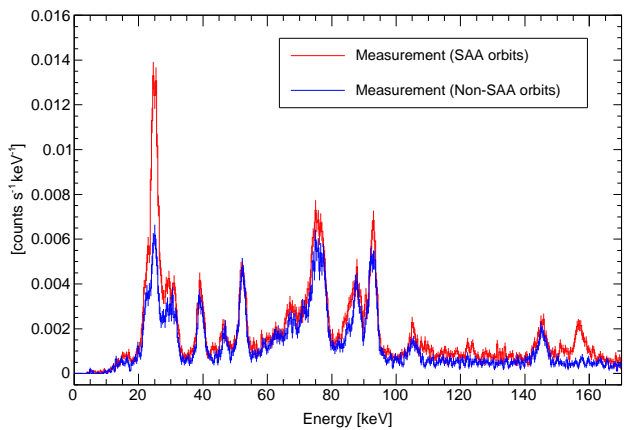}
\caption{Background spectra obtained with the CdTe detector of the HXI during the SAA orbits (red) and the non-SAA orbits (blue). Note that these spectra have the same normalization, showing remarkable similarity in their levels except for a few lines associated with isotopes of relatively short lifetimes.}
\label{fig:spectrum_measurement}
\end{center}
\end{figure}

We tested our simulation framework against in-orbit data obtained with the HXI \cite{Nakazawa:2016,Nakazawa:2017} onboard the \textit{Hitomi} \cite{Takahashi:2016, Takahashi:2017}.
The HXI was a focal plane imager composed of stacked Si and CdTe semiconductor detectors, covering an energy band from 5 keV to 80 keV.
Since this higher energy bound is limited by the mirror reflectivity, for unfocused photons and other particles, the CdTe detector layer was able to measure energy deposit up to $\sim$160 keV.
This instrument provided us with experimental data suitable for verifying the simulation since the satellite was put into a LEO with an altitude of 570 km and an inclination of 31$^\circ$, in which the detector suffered from a considerable amount of radioactivation caused by geo-magnetically trapped protons during passages through the South Atlantic Anomaly (SAA).
The orbital period was 96 minutes.

Figure~\ref{fig:hxi_drawing} shows a schematic drawing of the HXI in which only active detectors are shown (for passive 
materials of the supporting structure, see Figure~\ref{fig:mass_model}).
The main focal plane imager consists of four layers of double-sided strip detectors (DSDs) of Si and one layer of DSD of CdTe at the bottom \cite{Sato:2016}.
This imager unit is surrounded by thick $\mathrm{Bi_4Ge_3O_{12}}$ (BGO) scintillators for attenuation of protons that cause radioactivation as well as anti-coincidence of penetrating charged particles.
Each Si-DSD has a sensitive area of $32\times 32\;\mathrm{mm^2}$ and a thickness of 0.5\;mm.  The CdTe-DSD has the same sensitive area but has a thickness of 0.75 mm in order to have sufficient detection efficiency up to 80 keV.

As a material of high atomic numbers (Cd:48, Te:52), CdTe is very sensitive to hard X-ray photons, but the proton-induced radioactivation becomes a significant problem.
In this paper, we focus only on the background spectrum measured by the CdTe layer, since the Si layers do not suffer from radioactivation background.
The energy resolution of the CdTe-DSD is 2.0 keV (full width at half maximum) at 60 keV, which is sufficient to resolve lines arising from decays of radioactive isotopes.

The HXI system has a particle counter in close proximity to the main detector, to monitor charged particles that irradiate the main detector, which are mainly the geomagnetically trapped protons during the SAA passages or cosmic-ray protons outside the SAAs.
The particle monitor is a single avalanche photodiode (APD) identical to those used for 
readout of the BGO scintillator active shields.  This APD has a sensitive area of $10\times 10\;\mathrm{mm^2}$.
Figure~\ref{fig:particle_monitor} shows the time variability of the count rate of the particle monitor within an observation window of 48 hours.
The large peaks that have count rates of $\sim 10^2$--$10^3\;\mathrm{counts\;s^{-1}}$ correspond to 
the SAA passages.
Fluctuations with an amplitude of a few $\mathrm{counts\;s^{-1}}$ seen outside of the SAAs are due to variation of the geomagnetic cutoff rigidity.

The satellite orbits can be classified into SAA orbits and non-SAA orbits depending on whether the spacecraft passes 
through the SAA in a single 96-minute long orbit or not, as indicated in Figure~\ref{fig:particle_monitor}.
In the SAA orbits more isotopes of relatively short lifetimes than in the non-SAA orbits are present, so the background during the SAA orbits is higher. It is important to note that the main detector measured data only outside of the SAA passages---even during the SAA orbits---since its data acquisition had to be turned off in these high flux regions. The particle monitor, however, was always working.

We used spectra measured with the CdTe-DSD while the observatory was viewing the Earth.
Since \textit{Hitomi} was in LEO, its view was frequently hidden by the Earth, and data taken during Earth occultation can be regarded as pure background.
We selected events with an observatory elevation angle lower than $-5^\circ$ (the negative sign means looking down at the Earth).
We then eliminated events within the SAA passages with safety time margins of 251 s, and extracted spectra from the SAA orbits and the non-SAA orbits.
This selection condition can be written as
\begin{equation}
\label{eq:def_saa}
\begin{split}
&\text{SAA orbits:}\quad t_0> 251\;\mathrm{s}\quad\mathrm{and}\quad 251\;\mathrm{s}<t_1 < 5000\;\mathrm{s}, \\
&\text{Non-SAA orbits:}\quad t_0> 251\;\mathrm{s}\quad\mathrm{and}\quad 6000\;\mathrm{s}<t_1, \\
\end{split}
\end{equation}
where $t_0$ is time to the next SAA beginning and $t_1$ is time from the last SAA ending.
We applied these conditions to the Earth observation data provided via the \textit{Hitomi} archive, and extracted 
spectra with exposure times of 87.8 kilo-seconds for the SAA orbits and 70.0 kilo-seconds for the non-SAA orbits, where the standard event selection criteria and data reduction algorithms \cite{Hagino:2017} were applied.
Figure~\ref{fig:spectrum_measurement} shows the background spectra, making comparisons between the SAA orbits and the 
non-SAA orbits.  Both spectra are dominated by line features due to the in-orbit radioactivation with comparable 
intensities, though the SAA-orbit background displays higher intensities in some of the lines which originate from 
relatively short-lived isotopes.

\section{Simulation and Results}
\label{sec:results}

\begin{figure}[t]
\begin{center}
\includegraphics[width=9cm]{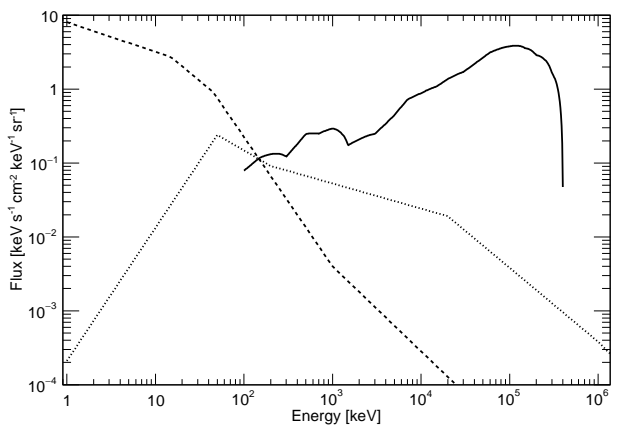}
\caption{Spectra of the particles/photons that produce background in the HXI-CdTe detector. The geomagnetically trapped protons associated with the SAA (solid), the CXB (dashed), and the albedo gamma rays (dotted) are shown.}
\label{fig:radiation_environment}
\end{center}
\end{figure}

\begin{table*}[t]
\caption{List of simulation sets}
\label{table:list_simulations}
\begin{center}
\begin{tabular}{cccc}
\hline\hline
Simulation set & Particle & Number & Time equivalent [$10^3$\;s] \\
\hline
Radioactivation Step 1 & trapped protons & $2.5\times 10^{10}$ & 24.6 (average rate) \\
Radioactivation Step 3 (in CdTe) & isotopes & & 320 \\
Radioactivation Step 3 (in BGOs) & isotopes & & 160 \\
Cosmic X-ray background & photon & $5.0\times 10^{11}$ & 310 \\
Albedo gamma rays & photon & $1.0\times 10^{11}$ & 1677 \\
\hline
\end{tabular}
\end{center}
\end{table*}%

This section describes the methods and conditions used in the simulations of the HXI background, their results, and comparisons with the measured data.
We used the \textit{ComptonSoft} framework, in which all simulations of photon and particle tracking and their interactions with matter are based on \textsc{Geant4}.
\textit{ComptonSoft} also treats calculation of charge transport inside the CdTe semiconductor detector that affects its detector response \cite{Hagino:2017, Odaka:2010}.
The mass model, shown in Figure~\ref{fig:mass_model}, and detector parameter database of the HXI system are identical to those used for generating the detector response released as an ``official'' product for astrophysical analysis \cite{Hagino:2017}.
This database includes properties of electron/hole transport in semiconductors, bias voltage, configuration of readout electrodes, and energy resolution and threshold energy in each readout channel of all the detectors including both the semiconductor imagers and the BGO active shields.
The mass model describes only the HXI system, not including the spacecraft structure.
This is appropriate since the HXI was mounted on an extensible optical bench, exposed to the radiation environment without any significant alteration by the spacecraft body.

\subsection{Radiation Environment}
\label{subsec:enviroment}

The background spectrum measured with the CdTe detector of the HXI consists of several components of different origins.
The main component is of radioactivation of the CdTe detector material by the SAA protons, which produce internal instrumental background that is extremely difficult to eliminate.
Radioactivation of external heavy material can also be an issue, so we consider activation of the surrounding BGO shields, which are the very heaviest parts around the main imager.
Background X-ray and gamma-ray photons, mainly cosmic X-ray background (CXB) and albedo gamma rays, coming through openings of the shields and the camera baffles also make significant contributions to the background spectrum.
In addition, there is internal background due to naturally contaminated radioactive isotopes in materials of the detector system.
This component was observed in ground measurements.
Although other backgrounds originating from cosmic-ray protons, electrons and positrons, and albedo neutrons, for example, are possible, we confirmed by Monte Carlo simulation that these contributions are negligible, i.e., at most two orders of magnitude lower than the measured level.
The majority of the background spectrum can be explained by radioactivation in the CdTe detector and the BGO active shields induced by the trapped protons, the photon background (CXB and albedo), and the internal background due to the natural contamination.

The populations of primaries to be simulated in this work are the CXB, albedo gamma rays, and geomagnetically trapped protons associated with the SAA.
These spectra are shown in Figure~\ref{fig:radiation_environment}.
Table~\ref{table:list_simulations} also lists the simulation components.
The spectra of the CXB and albedo gamma rays were adopted from our previous work \cite{Mizuno:2010}.
We generated the spectrum of the trapped protons via \textit{SPENVIS} (Space Environment Information System)\footnote{https://www.spenvis.oma.be} by inputting information on the orbit of \textit{Hitomi}, and assuming the AP-8 model \cite{Gaffey:1994} at solar minimum.
Note that this spectrum is time-averaged while its flux is highly time variable along the orbit.

In all simulations, primary particles were generated so that they had homogeneous isotropic distributions in a sphere of radius 51 cm which contained the entire HXI system.
The direction of each particle was isotropically sampled from a solid angle of 4$\pi$.
The total number of primaries are shown in Table~\ref{table:list_simulations}.
While the CXB and the albedo gamma rays do not have isotropic distributions, these angular distributions are treated at a later stage (\S\ref{subsec:photon_background}) by a weighting method since they depend on the spacecraft attitude.
Though the trapped protons also do not have an isotropic distribution, our simplification of the isotropic distribution probably works well for generating activated isotopes.

\subsection{Proton-induced Radioactivation}
\label{subsec:radioactivation}

\begin{figure*}[t]
\begin{center}
\includegraphics[width=9cm]{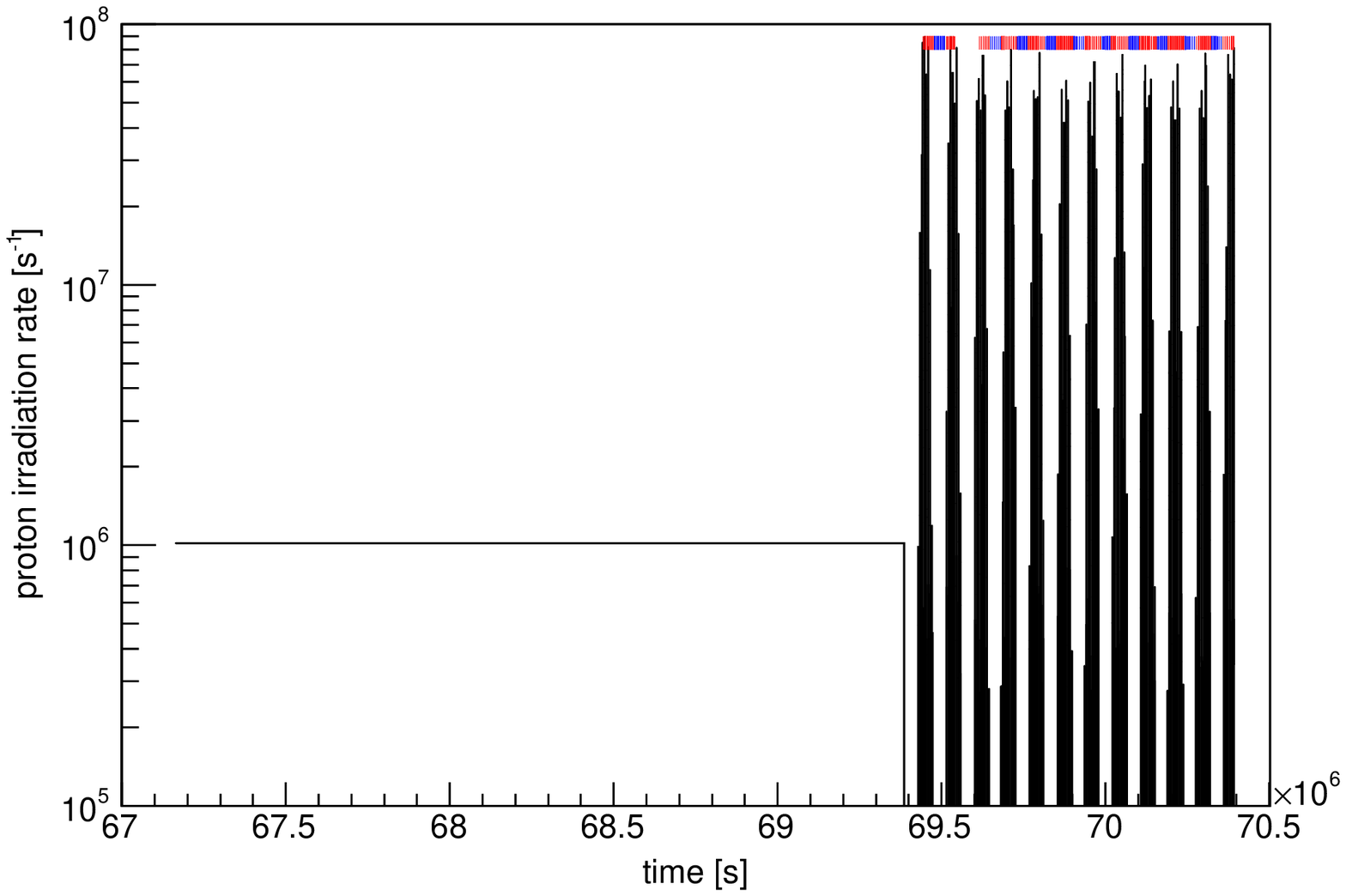}
\includegraphics[width=9cm]{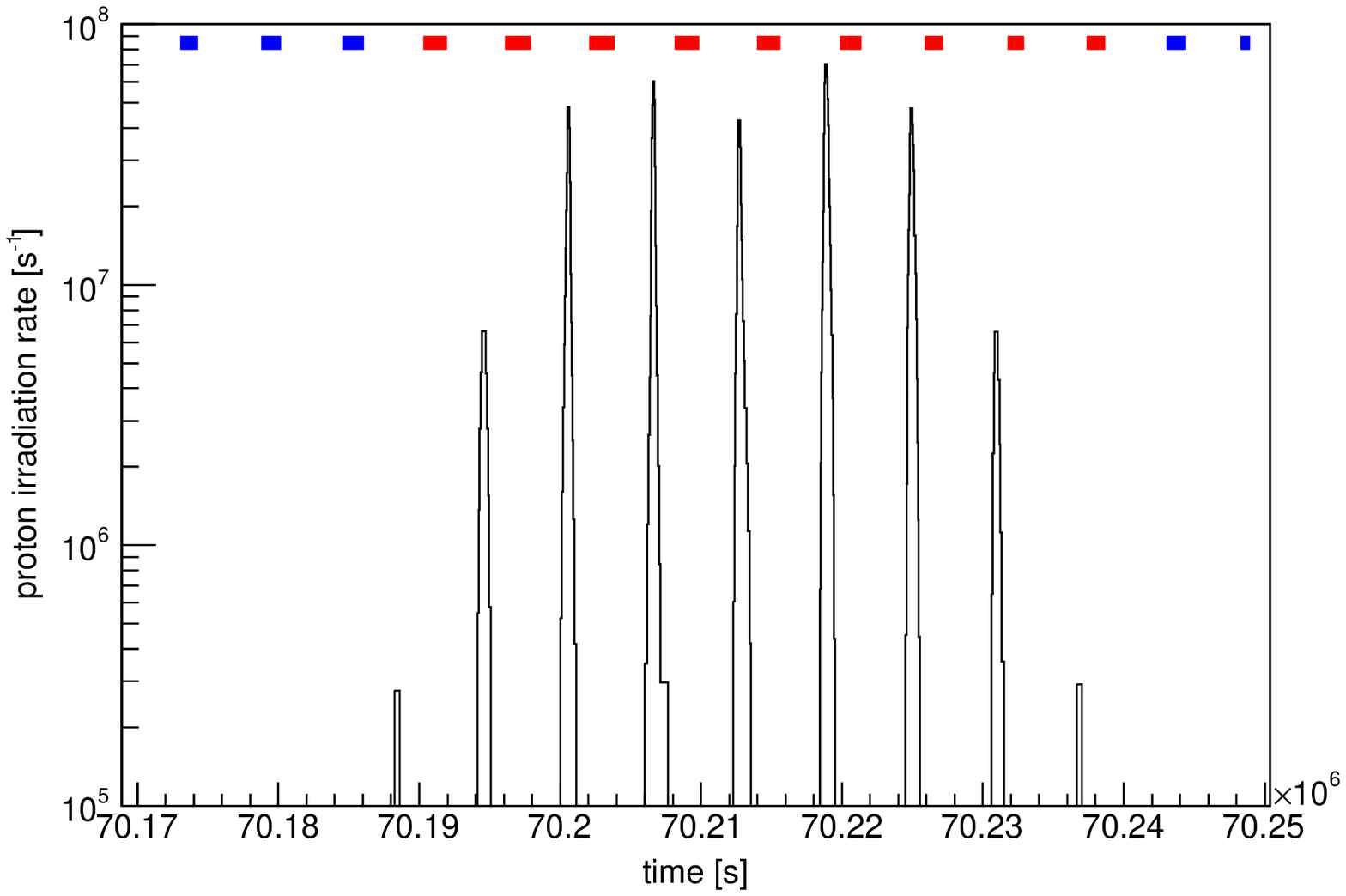}
\caption{Time profile of the irradiation rate of the trapped protons assumed in Step 2 of the radioactivation simulation. The measurement time windows are also indicated for the SAA orbits (red) and the non-SAA orbits (blue). The left panel shows the entire history while the right panel focuses on SAA passages in one day.}
\label{fig:time_profile}
\end{center}
\end{figure*}

\begin{figure}[t]
\begin{center}
\includegraphics[width=9cm]{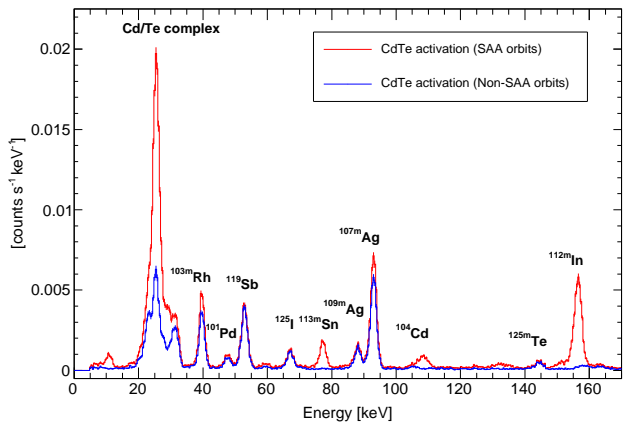}
\caption{Simulated spectra of radioactivation background in CdTe during the SAA orbits (red) and the non-SAA orbits (blue). 
         Isotopes that emit strong lines are indicated.}
\label{fig:spectrum_activation}
\end{center}
\end{figure}

Simulation of radioactivation was performed in three steps as described in \S\ref{sec:method}.
Step 1 treated the geomagnetically trapped protons described in \S\ref{subsec:enviroment} as primary particles, simulating $2.5\times 10^{10}$ Monte Carlo events.
We used the same physics settings (i.e., the binary-cascade hadron physics) adopted in \S\ref{sec:verification}.
The output of this step, which is a list of generated isotopes and their number, was given to the next 
step.

Step 2 considered the history of the particle irradiation since the launch of the satellite, and the measurement time windows.
Figure~\ref{fig:time_profile} shows the time profile of the irradiation rate assumed for this calculation as well as the measurement windows.
The average rate of the irradiation was $1.016\times 10^{6}\;\mathrm{particles\;s^{-1}}$ or 
$9.9\;\mathrm{particles\;s^{-1}\;cm^{-2}\;sr^{-1}}$ in the entire energy band according to the \textit{SPENVIS} model (\S\ref{subsec:enviroment}).
We used the relative amplitude of the proton irradiation based on the particle monitor measurement with normalization scaled to the averaged rate.
To reduce the number of time sections (labeled by $m$) in Equation~(\ref{eq:sum_time_profile}), we applied the Bayesian block algorithm \cite{Scargle:2013a, Scargle:2013b} to the raw monitor data, 
yielding an even smaller number of time sections that have statistically significant differences in the irradiation rate.
For the period before the HXI system (and its data acquisition) was turned on, in which the particle monitor data were not available, we assumed for simplicity a constant irradiation at the average rate.
We performed Step 2 calculations separately for the measurements in the SAA and non-SAA orbits, obtaining two sets of isotope decay rate lists as inputs for the next step.

Step 3 conducted decay simulations for all the isotopes and summed the results over all different isotopes.
We show spectra obtained by these simulations for both the SAA orbits and non-SAA orbits in 
Figure~\ref{fig:spectrum_activation}.  Both those spectra consist of line features and underlying continua while the 
non-SAA background is clearly lower particularly at lines associated with short-lived isotopes.
Isotopes significantly responsible for the radioactivation background are listed in Table~\ref{table:isotopes_saa} (SAA) 
and Table~\ref{table:isotopes_nonsaa} (non-SAA), in which we select isotopes that produce background rates higher than 
$5\times 10^{-4}\;\mathrm{counts\;s^{-1}}$ within an energy range of 15--165 keV.
Most of the line features, also shown in Tables~\ref{table:isotopes_saa} and \ref{table:isotopes_nonsaa}, come from 
isomeric transitions (ITs) of meta-stable nuclei (so-called isomers) or daughter isotopes of $\beta^\pm$ decays or electron 
capture (EC).
The complex feature seen around 20--30 keV is composed of atomic transition lines associated with EC of isotopes activated from Cd and Te.
The continuum is produced mainly by isotopes $^{127}\mathrm{Te}$, $^{129}\mathrm{Te}$, and $^{112}\mathrm{In}$ during the SAA orbits.
All of these three isotopes decay by emitting $\beta^-$ while $^{112}\mathrm{In}$ also has electron capture (EC) channels.
Importantly, $^{127}\mathrm{Te}$ is the only dominant source in the radioactivation continuum during the non-SAA orbits 
since the other two decay even faster.
Comparison to the data of the simulation including all components will be discussed in \S\ref{subsec:comparison}.

We also performed Step 3 for BGO shields in the same manner as the CdTe activation.
We discuss its results in \S\ref{subsec:comparison} together with other background components as it does not make  
significant contributions to line features except for the atomic transition complex at 68--87 keV due to bismuth activation.

\begin{table*}
\caption{Isotopes largely responsible for background in the SAA orbits}
\label{table:isotopes_saa}
\begin{center}
\begin{threeparttable}
\begin{tabular}{cccccccc}
\hline\hline
Isotope & Z\tnote{a} & A\tnote{b} & E\tnote{c} [keV] & Halflife [s] & Decay rate\tnote{d} [s$^{-1}$] & Count rate\tnote{e} [s$^{-1}$] & Line features\tnote{f} [keV] \\
\hline
$^{107m}\mathrm{Ag}$ & 47 & 107 & 93.1 & 4.43e$+$01 & 2.14e$-$02 & 1.76e$-$02 & 93 \\
$^{112m}\mathrm{In}$ & 49 & 112 & 156.6 & 1.24e$+$03 & 2.34e$-$02 & 1.74e$-$02 & 157 \\
$^{105m}\mathrm{Ag}$ & 47 & 105 & 25.5 & 4.34e$+$02 & 1.99e$-$02 & 1.69e$-$02 & 25 \\
$^{107}\mathrm{Cd}$ & 48 & 107 & 0.0 & 2.34e$+$04 & 2.14e$-$02 & 1.46e$-$02 & 25 \\
$^{119}\mathrm{Sb}$ & 51 & 119 & 0.0 & 1.37e$+$05 & 1.48e$-$02 & 1.21e$-$02 & 53, 28 \\
$^{112}\mathrm{In}$ & 49 & 112 & 0.0 & 8.93e$+$02 & 3.07e$-$02 & 1.15e$-$02 & 27, continuum \\
$^{103m}\mathrm{Rh}$ & 45 & 103 & 39.8 & 3.37e$+$03 & 1.08e$-$02 & 9.22e$-$03 & 40 \\
$^{127}\mathrm{Te}$ & 52 & 127 & 0.0 & 3.37e$+$04 & 2.76e$-$02 & 8.98e$-$03 & continuum \\
$^{103}\mathrm{Pd}$ & 46 & 103 & 0.0 & 1.47e$+$06 & 8.20e$-$03 & 5.72e$-$03 & 23 \\
$^{120}\mathrm{Sb}$ & 51 & 120 & 0.0 & 9.53e$+$02 & 1.20e$-$02 & 4.59e$-$03 & 29 \\
$^{113m}\mathrm{Sn}$ & 50 & 113 & 77.4 & 1.28e$+$03 & 5.43e$-$03 & 4.48e$-$03 & 77 \\
$^{129}\mathrm{Te}$ & 52 & 129 & 0.0 & 4.18e$+$03 & 3.55e$-$02 & 4.44e$-$03 & continuum \\
$^{105}\mathrm{Cd}$ & 48 & 105 & 0.0 & 3.33e$+$03 & 1.71e$-$02 & 4.25e$-$03 & 25 \\
$^{106}\mathrm{Ag}$ & 47 & 106 & 0.0 & 1.44e$+$03 & 1.61e$-$02 & 3.72e$-$03 & 24 \\
$^{125}\mathrm{I}$ & 53 & 125 & 0.0 & 5.13e$+$06 & 4.45e$-$03 & 3.57e$-$03 & 67, 40 \\
$^{104}\mathrm{Cd}$ & 48 & 104 & 0.0 & 3.46e$+$03 & 5.83e$-$03 & 2.72e$-$03 & 109 \\
$^{124}\mathrm{I}$ & 53 & 124 & 0.0 & 3.61e$+$05 & 1.34e$-$02 & 2.46e$-$03 & 32 \\
$^{101}\mathrm{Pd}$ & 46 & 101 & 0.0 & 3.05e$+$04 & 4.96e$-$03 & 2.38e$-$03 & 48 \\
$^{118}\mathrm{Te}$ & 52 & 118 & 0.0 & 5.18e$+$05 & 3.52e$-$03 & 2.31e$-$03 & 30 \\
$^{128}\mathrm{I}$ & 53 & 128 & 0.0 & 1.50e$+$03 & 1.78e$-$02 & 2.28e$-$03 & 32, continuum \\
$^{124m}\mathrm{Sb}$ & 51 & 124 & 36.8 & 1.21e$+$03 & 2.58e$-$03 & 2.21e$-$03 & 26 \\
$^{126}\mathrm{I}$ & 53 & 126 & 0.0 & 1.12e$+$06 & 1.19e$-$02 & 2.16e$-$03 & 32, continuum \\
$^{109m}\mathrm{Ag}$ & 47 & 109 & 88.0 & 3.96e$+$01 & 2.54e$-$03 & 2.12e$-$03 & 88 \\
$^{123}\mathrm{I}$ & 53 & 123 & 0.0 & 4.76e$+$04 & 1.41e$-$02 & 1.61e$-$03 & 159, (191), (163), 32 \\
$^{115}\mathrm{Cd}$ & 48 & 115 & 0.0 & 1.92e$+$05 & 1.07e$-$02 & 1.42e$-$03 & continuum \\
$^{125m}\mathrm{Te}$ & 52 & 125 & 144.8 & 4.96e$+$06 & 1.69e$-$03 & 1.34e$-$03 & 145 \\
$^{109}\mathrm{Cd}$ & 48 & 109 & 0.0 & 3.99e$+$07 & 1.96e$-$03 & 1.25e$-$03 & 25 \\
$^{111}\mathrm{Sn}$ & 50 & 111 & 0.0 & 2.12e$+$03 & 2.25e$-$03 & 9.16e$-$04 & 28 \\
$^{118}\mathrm{Sb}$ & 51 & 118 & 0.0 & 2.16e$+$02 & 5.25e$-$03 & 8.75e$-$04 & 29 \\
$^{127m}\mathrm{Te}$ & 52 & 127 & 88.2 & 9.17e$+$06 & 1.06e$-$03 & 8.69e$-$04 & 88 \\
$^{100m}\mathrm{Rh}$ & 45 & 100 & 107.6 & 2.76e$+$02 & 1.03e$-$03 & 8.41e$-$04 & 108 \\
$^{117}\mathrm{Sb}$ & 51 & 117 & 0.0 & 1.01e$+$04 & 7.26e$-$03 & 8.38e$-$04 & (188), 29 \\
$^{111m}\mathrm{Cd}$ & 48 & 111 & 396.2 & 2.91e$+$03 & 2.86e$-$02 & 8.28e$-$04 & (396), 150, (245) \\
$^{103}\mathrm{Ag}$ & 47 & 103 & 0.0 & 3.94e$+$03 & 8.88e$-$03 & 8.19e$-$04 & (291), 24 \\
$^{109}\mathrm{In}$ & 49 & 109 & 0.0 & 1.50e$+$04 & 1.62e$-$02 & 7.19e$-$04 & (230), 27 \\
$^{119}\mathrm{Te}$ & 52 & 119 & 0.0 & 5.78e$+$04 & 7.65e$-$03 & 6.72e$-$04 & 30 \\
$^{100}\mathrm{Pd}$ & 46 & 100 & 0.0 & 3.14e$+$05 & 2.08e$-$03 & 5.97e$-$04 & (182), 23 \\
$^{109m}\mathrm{Cd}$ & 48 & 109 & 59.5 & 1.20e$-$05 & 6.92e$-$04 & 5.66e$-$04 & 59 \\
$^{101m}\mathrm{Rh}$ & 45 & 101 & 157.3 & 3.75e$+$05 & 6.67e$-$03 & 5.47e$-$04 & 157, 22, (307), (329) \\
$^{121}\mathrm{Sn}$ & 50 & 121 & 0.0 & 9.73e$+$04 & 1.01e$-$03 & 5.41e$-$04 & continuum \\
$^{129m}\mathrm{Te}$ & 52 & 129 & 105.5 & 2.90e$+$06 & 1.01e$-$03 & 5.41e$-$04 & 106 \\
$^{111}\mathrm{Ag}$ & 47 & 111 & 0.0 & 6.44e$+$05 & 2.55e$-$03 & 5.34e$-$04 & continuum \\
$^{116}\mathrm{Sb}$ & 51 & 116 & 0.0 & 9.48e$+$02 & 8.11e$-$03 & 5.34e$-$04 & 29 \\
\hline
\end{tabular}
\begin{tablenotes}
\item[a] Atomic number of isotope.
\item[b] Atomic mass number of isotope.
\item[c] Excited energy of isotope. A value of 0.0 indicates the ground state of nucleus.
\item[d] Decay rates within the CdTe detector.
\item[e] Measurement count rates within the CdTe detector.	
\item[f] Observable line features produced by the isotope in order of intensity. Energies in parentheses are outside of 
         the measurable range of the detector.
\end{tablenotes}
\end{threeparttable}
\end{center}
\end{table*}%

\begin{table*}
\caption{Isotopes largely responsible for background in the non-SAA orbits}
\label{table:isotopes_nonsaa}
\begin{center}
\begin{threeparttable}
\begin{tabular}{cccccccc}
\hline\hline
Isotope & Z & A & E [keV] & Halflife [s] & Decay rate [s$^{-1}$] & Count rate [s$^{-1}$] & Line features [keV] \\
\hline
$^{107m}\mathrm{Ag}$ & 47 & 107 & 93.1 & 4.43e$+$01 & 1.77e$-$02 & 1.45e$-$02 & 93 \\
$^{119}\mathrm{Sb}$ & 51 & 119 & 0.0 & 1.37e$+$05 & 1.50e$-$02 & 1.23e$-$02 & 53, 28 \\
$^{107}\mathrm{Cd}$ & 48 & 107 & 0.0 & 2.34e$+$04 & 1.76e$-$02 & 1.21e$-$02 & 25 \\
$^{127}\mathrm{Te}$ & 52 & 127 & 0.0 & 3.37e$+$04 & 2.54e$-$02 & 8.28e$-$03 & continuum \\
$^{103m}\mathrm{Rh}$ & 45 & 103 & 39.8 & 3.37e$+$03 & 8.34e$-$03 & 7.15e$-$03 & 40 \\
$^{103}\mathrm{Pd}$ & 46 & 103 & 0.0 & 1.47e$+$06 & 8.19e$-$03 & 5.72e$-$03 & 23 \\
$^{125}\mathrm{I}$ & 53 & 125 & 0.0 & 5.13e$+$06 & 4.42e$-$03 & 3.53e$-$03 & 67, 40 \\
$^{124}\mathrm{I}$ & 53 & 124 & 0.0 & 3.61e$+$05 & 1.35e$-$02 & 2.47e$-$03 & 32 \\
$^{118}\mathrm{Te}$ & 52 & 118 & 0.0 & 5.18e$+$05 & 3.53e$-$03 & 2.31e$-$03 & 30 \\
$^{126}\mathrm{I}$ & 53 & 126 & 0.0 & 1.12e$+$06 & 1.19e$-$02 & 2.16e$-$03 & 32, continuum \\
$^{101}\mathrm{Pd}$ & 46 & 101 & 0.0 & 3.05e$+$04 & 4.40e$-$03 & 2.12e$-$03 & 48 \\
$^{109m}\mathrm{Ag}$ & 47 & 109 & 88.0 & 3.96e$+$01 & 2.51e$-$03 & 2.09e$-$03 & 88 \\
$^{123}\mathrm{I}$ & 53 & 123 & 0.0 & 4.76e$+$04 & 1.36e$-$02 & 1.56e$-$03 & 159, (191), (163), 32 \\
$^{115}\mathrm{Cd}$ & 48 & 115 & 0.0 & 1.92e$+$05 & 1.08e$-$02 & 1.42e$-$03 & continuum \\
$^{125m}\mathrm{Te}$ & 52 & 125 & 144.8 & 4.96e$+$06 & 1.67e$-$03 & 1.33e$-$03 & 145 \\
$^{109}\mathrm{Cd}$ & 48 & 109 & 0.0 & 3.99e$+$07 & 1.94e$-$03 & 1.23e$-$03 & 25 \\
$^{127m}\mathrm{Te}$ & 52 & 127 & 88.2 & 9.17e$+$06 & 1.05e$-$03 & 8.59e$-$04 & 88 \\
$^{119}\mathrm{Te}$ & 52 & 119 & 0.0 & 5.78e$+$04 & 7.57e$-$03 & 6.62e$-$04 & 30 \\
$^{118}\mathrm{Sb}$ & 51 & 118 & 0.0 & 2.16e$+$02 & 3.53e$-$03 & 6.12e$-$04 & 29 \\
$^{100}\mathrm{Pd}$ & 46 & 100 & 0.0 & 3.14e$+$05 & 2.09e$-$03 & 6.03e$-$04 & (182), 23 \\
$^{101m}\mathrm{Rh}$ & 45 & 101 & 157.3 & 3.75e$+$05 & 6.75e$-$03 & 5.56e$-$04 & 157, 22, (307), (329) \\
$^{129m}\mathrm{Te}$ & 52 & 129 & 105.5 & 2.90e$+$06 & 1.00e$-$03 & 5.41e$-$04 & 106 \\
$^{121}\mathrm{Sn}$ & 50 & 121 & 0.0 & 9.73e$+$04 & 1.01e$-$03 & 5.41e$-$04 & continuum \\
$^{111}\mathrm{Ag}$ & 47 & 111 & 0.0 & 6.44e$+$05 & 2.56e$-$03 & 5.34e$-$04 & continuum \\
\hline
\end{tabular}
\begin{tablenotes}
\item[] For description of columns, see Table~\ref{table:isotopes_saa}.
\end{tablenotes}
\end{threeparttable}
\end{center}
\end{table*}%

\subsection{Photon backgrounds}
\label{subsec:photon_background}

\begin{figure*}[t]
\begin{center}
\includegraphics[width=9cm]{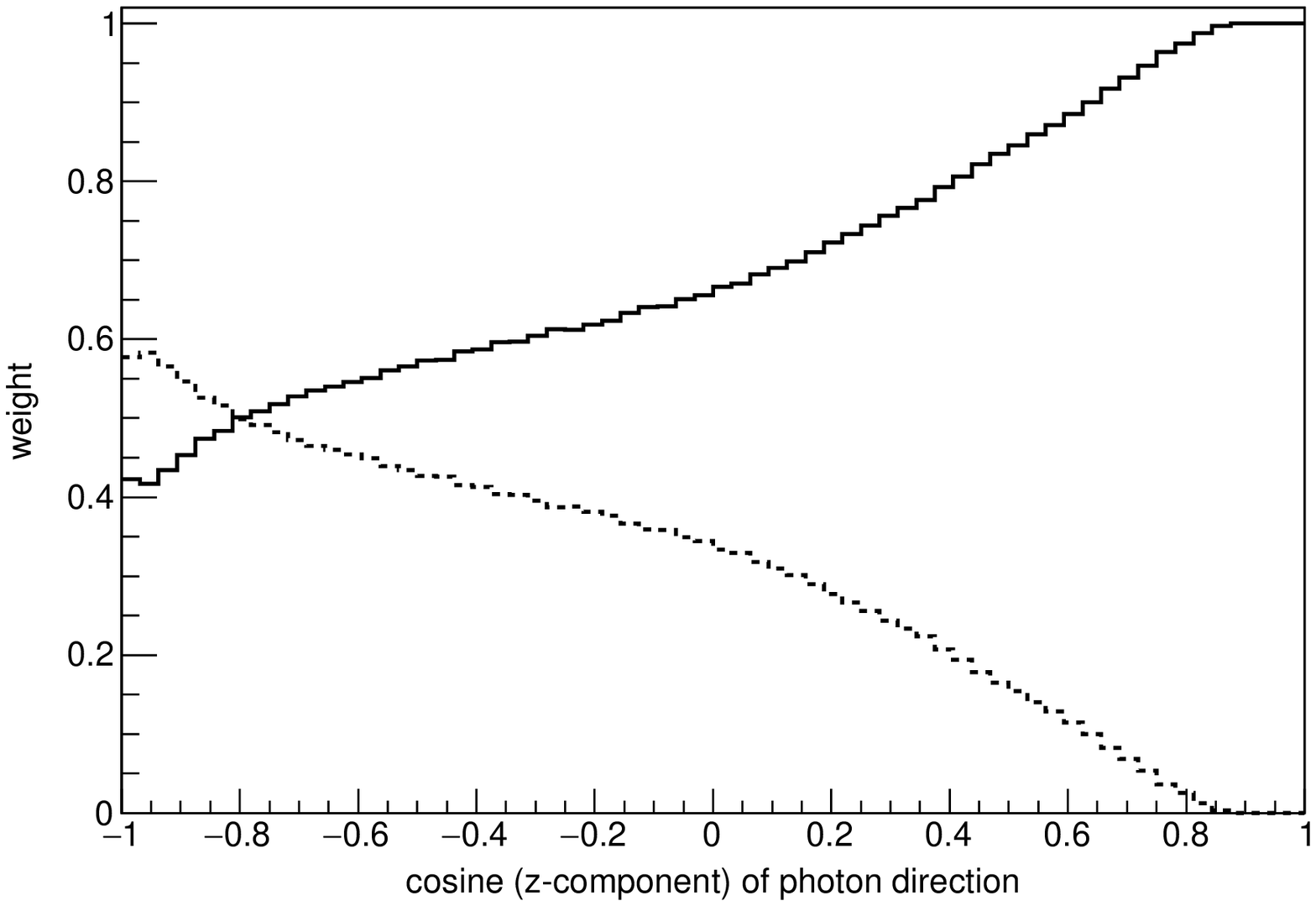}
\includegraphics[width=9cm]{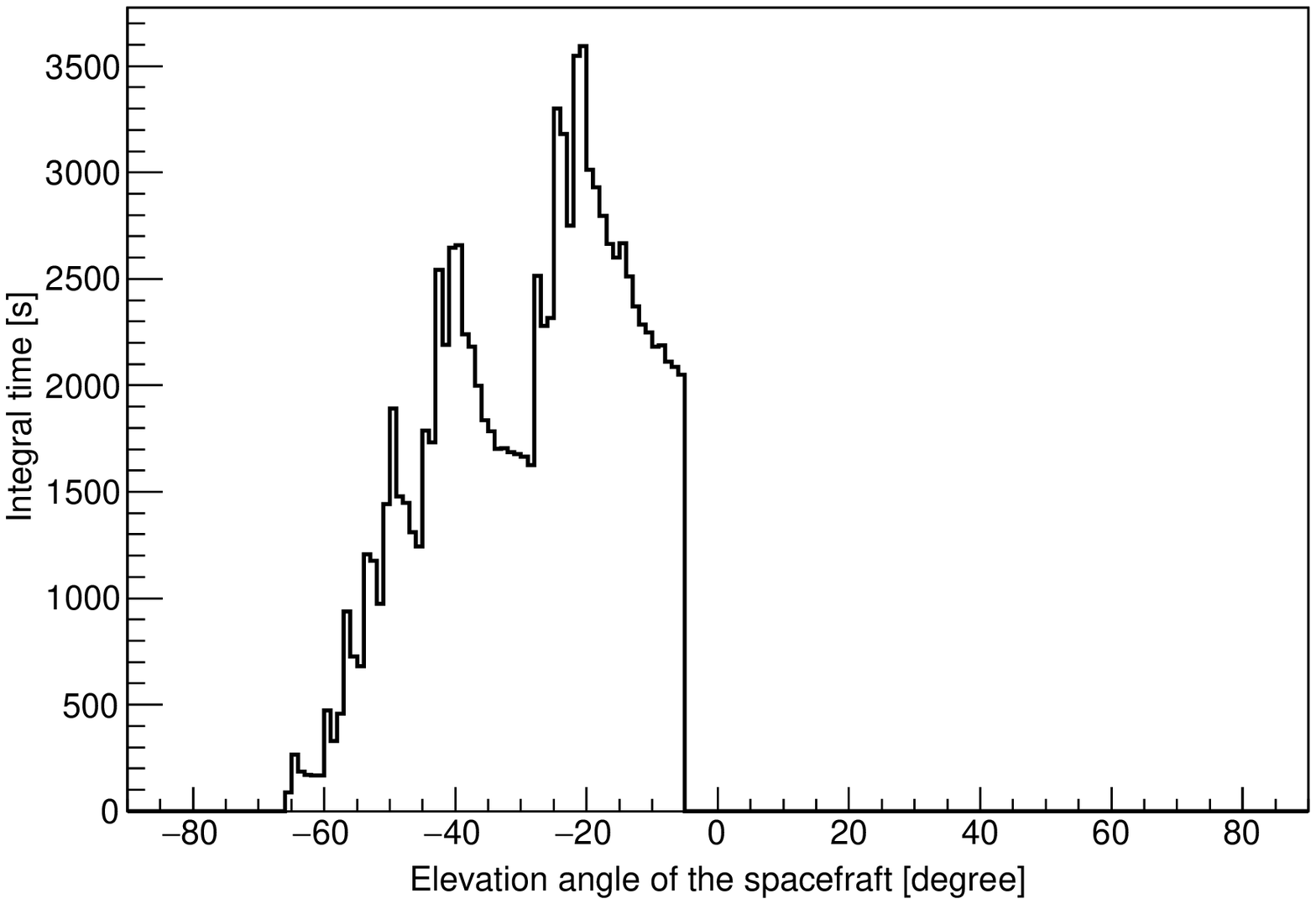}
\caption{Left: angular dependence of the CXB (solid) and the albedo gamma rays (dashed) with respect to the HXI optical axis for the measurement during the SAA orbits. Right: distribution of the elevation angle of the spacecraft during the SAA orbits. }
\label{fig:angular_distribution_photons}
\end{center}
\end{figure*}

We performed two sets of photon background simulations for the CXB and albedo gamma rays.
Attenuation of the photon flux by the spacecraft body, which is not included in the simulation mass model, was assumed
to be negligible as well as that of protons.
The only exception to this occurs in photon paths immediately outside of the field of view since CXB shields 
made of lead or tin were placed on the satellite plate structures in order to suppress photon leakage through the 
aperture of the camera baffle.
This type of photon leakage would cause a significant background, as reported in \textit{NuSTAR} \cite{Wik:2014}.
To consider the shield attenuation, we processed the event lists output from the simulations, weighting each event by 
an attenuation factor calculated by the photo-absorption cross section and effective thickness of the shield for its 
photon path.

We assumed that the CXB photons come from the sky (elevation angle $> -23^\circ.6$) while the albedo gamma rays come from the Earth (elevation angle $< -23^\circ.6$), taking account of the satellite altitude of 570\;km.
The left panel of Figure~\ref{fig:angular_distribution_photons} shows angular distributions of the two photon backgrounds with respect to the spacecraft (the HXI optical axis) which were used for weighting each event in order to extract the background spectra that considered the spacecraft's attitude to the sky and the Earth.
These angular dependences relative to the spacecraft were obtained by averaging the distributions over the elevation angle distribution along the orbit, which is shown in the right panel of Figure~\ref{fig:angular_distribution_photons}.

\subsection{Comparison to Measured Data}
\label{subsec:comparison}

\begin{figure*}[t]
\begin{center}
\includegraphics[width=18cm]{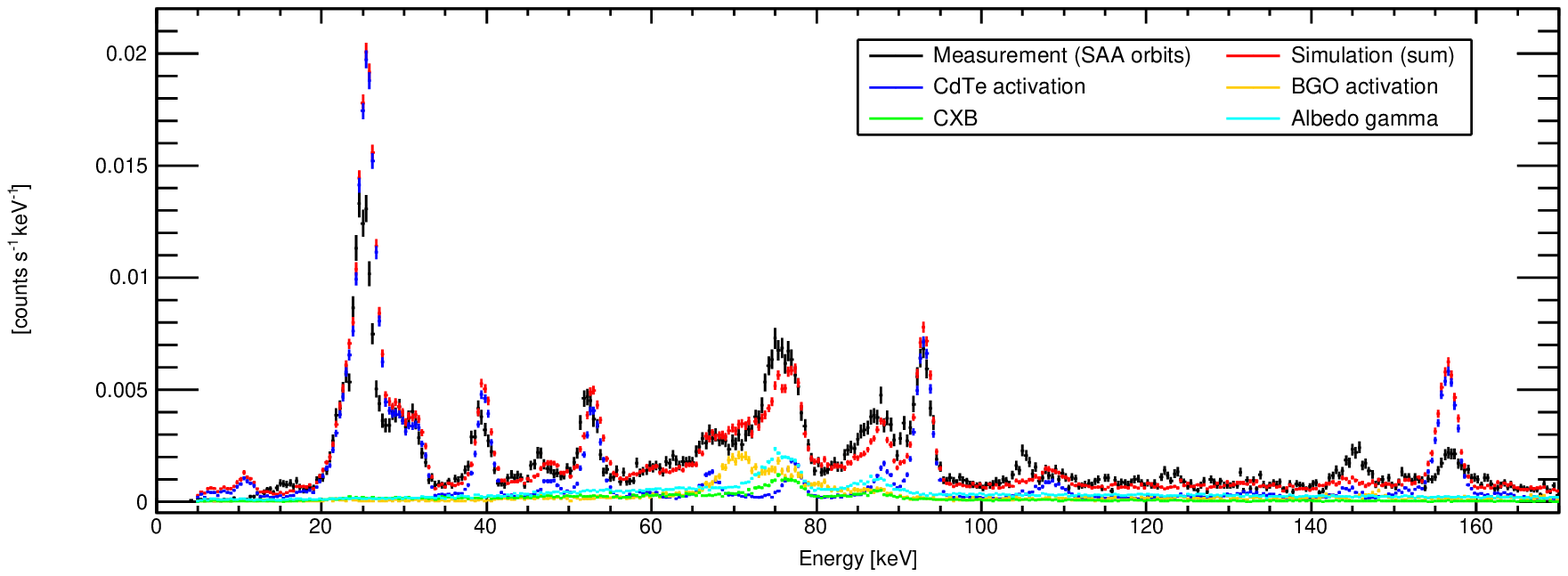}
\includegraphics[width=18cm]{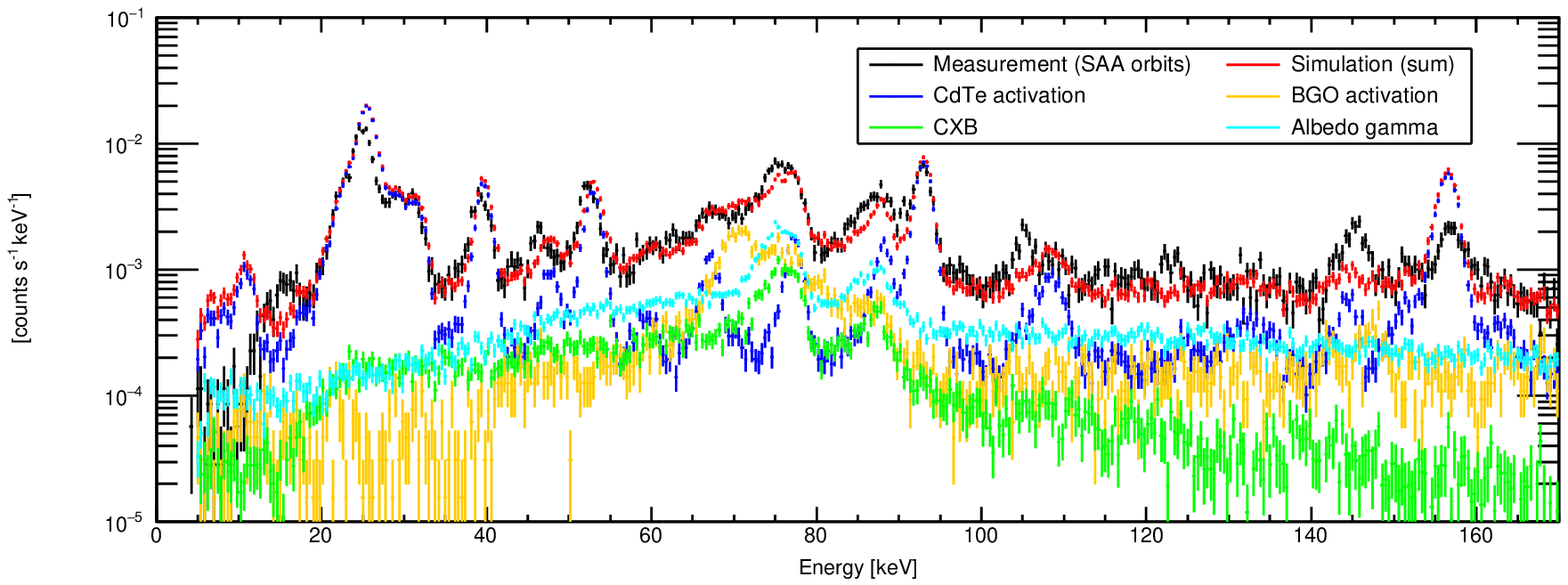}
\caption{Background spectrum (black) of the HXI CdTe-DSD measured during the SAA orbits compared with simulation (red). All simulation components are also shown: radioactivation in CdTe (blue), radioactivation in BGO (yellow), CXB (green), and albedo gamma rays (cyan). The top panel is shown in linear scale, while the bottom is logarithmic. Although the data points are shown in the range of 5--170 keV, efficiency of readout rapidly declines below $\sim$10 keV due to trigger thresholds and above $\sim$160 keV due to limited analog-to-digital converter ranges (see \S\ref{sec:discussion}).}
\label{fig:background_spectra_saa}
\end{center}
\end{figure*}

\begin{figure*}[t]
\begin{center}
\includegraphics[width=18cm]{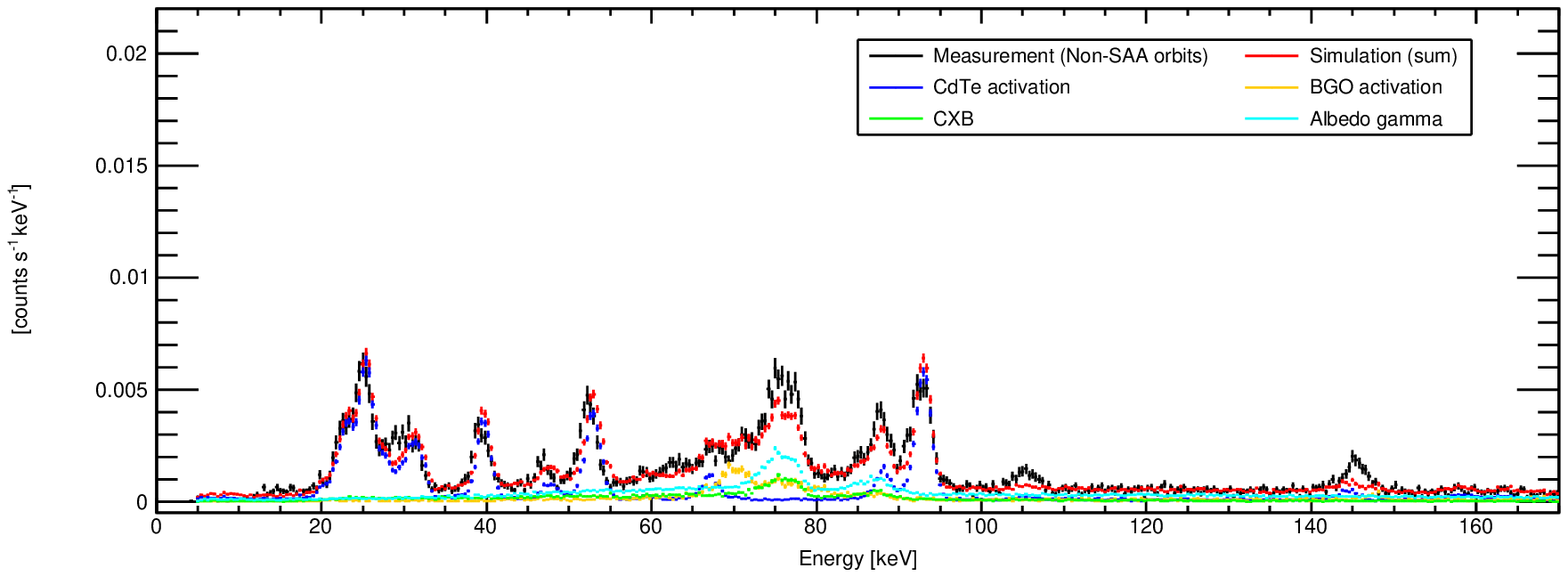}
\includegraphics[width=18cm]{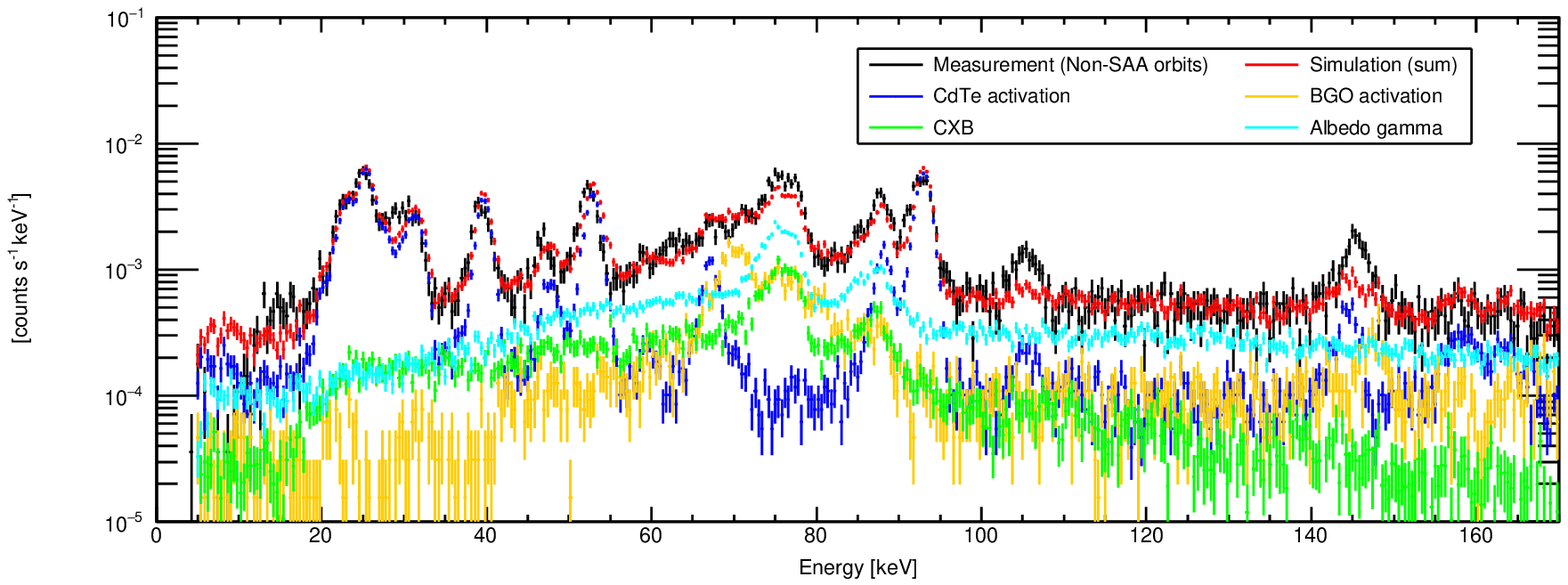}
\caption{Same as Figure~\ref{fig:background_spectra_saa} but for the non-SAA orbits.}
\label{fig:background_spectra_nonsaa}
\end{center}
\end{figure*}


In order to compare the simulations to measured data, we summed the simulations of all possible background origins, i.e., 
radioactivation by the trapped protons in CdTe and BGO, the CXB, and the albedo gamma rays.
Prior to the summation, we scaled the albedo gamma-ray component by a factor of 1.5 to account for residuals seen in the 
continuum above 100 keV and in the fluorescence lines from bismuth (Bi in the BGO shields) at 75--77 keV (K$\alpha$) 
and 87 keV (K$\beta$), while no additional scaling was applied to the other components.
This scaling was necessary probably due to the large uncertainties in the albedo gamma-ray flux.
Nevertheless, residuals at the Bi fluorescence remain; this point will be further discussed in \S\ref{sec:discussion}.
The results are shown in Figure~\ref{fig:background_spectra_saa} and Figure~\ref{fig:background_spectra_nonsaa} for the 
SAA and non-SAA orbits, respectively.

The simulation results show remarkably successful agreement with the measured spectra.
The background spectra consist of line features and continuum.
The radioactivation of the CdTe detector produces most of the lines observed in the background spectra while the photon backgrounds are dominantly responsible for bismuth fluorescence lines at 75--77 keV (K$\alpha$) and 87 keV (K$\beta$).
The radioactivation of the BGO shields also produces the bismuth fluorescence as well as atomic deexcitation lines of the secondary isotopes of bismuth activation, which are distributed in an energy range of 68--87 keV.
A few observed lines are missing (or are too weak) in the simulation though some of them can probably be attributed to natural contamination from radioactive isotopes as discussed in the next section.

In contrast, the continuum consists of multiple components, and the photon backgrounds have more significant contributions.
The albedo gamma-ray component constitutes a large fraction therein as its hard spectrum (see Fig.~\ref{fig:radiation_environment}) has large penetrating power though the shields, generating significant scattered photons in the vicinity of the detector.
However, the radioactivation in CdTe also contributes to the continuum.
It should be stressed that the CdTe radioactivation produces the continuum emission only via $^{127}\mathrm{Te}$, $^{129}\mathrm{Te}$, and $^{112}\mathrm{In}$, as already mentioned in \S\ref{subsec:radioactivation}, since this fact is useful for modeling of the time variability of the background.

We observe slight energy shifts between the measured spectrum and the simulation of the CdTe radioactivation.
Since the upper bound of the ``official'' energy range of the HXI for astronomical observation is 80 keV, a correction function of energy gain that converts measured pulse heights to energy deposits is valid only below 80 keV, and therefore the gain correction function can be distorted above this energy.
Even in the valid energy range, however, the simulation line energies are systematically higher than the measured peak energies by $\sim$0.5 keV.
These differences are clearly seen in two lines at 39.8 keV ($^{103m}\mathrm{Rh}$) and 53.0 keV ($^{119}\mathrm{Sb}$).
We ascribe these differences in the line energies to the response of the CdTe detector used in the simulations that describes the charge collection efficiency in the semiconductor as a function of position of energy deposit \cite{Odaka:2010, Hagino:2012}.
Parameters of the detector response of the CdTe-DSD were optimized for astronomical observation by using calibration measurements of gamma-ray sources \cite{Hagino:2017}.
The response to the external irradiation by the gamma-ray sources can be different from that of the radioactivation signals generated inside the detector wafer.

\section{Discussion}
\label{sec:discussion}

As shown in the previous section, the full simulation results show very good agreement with the in-orbit spectra obtained with the HXI.
In this section, we first discuss important factors that are relevant to the accuracy of the simulation.
However, we find that some disagreement still remains, and we discuss possible origins of the discrepancy.
We also suggest possible improvements in the model accuracy for future hard X-ray missions.

One of the most important improvements regarding physical processes in this work is full treatment of isomeric (meta-stable) levels of isotopes.
In general, inelastic interactions of protons produce many isomers, resulting in line backgrounds via isomeric transitions (ITs).
There are actually many IT lines seen in the measured background spectra of the CdTe sensor.
In the simulation framework, these ITs can be divided into two categories: (1) direct generation of isomers via the inelastic processes of primary protons, and (2) radioactive decays to isomeric levels from other activated isotopes (at their ground states in many cases).
The former case was newly introduced in \textsc{Geant4} version 10.03, while the latter has been processed by a conventional scheme of radioactive decay.
Since the treatment of isomeric levels in the inelastic interactions is significantly complicated, they had been forced to decay to their ground states before version 10.03.

Our simulation framework can handle multiple isomeric levels in each isotope in all simulation steps---from the secondary isotope generations to their decays.
The nuclear level information is provided by the \texttt{Photon\-Evaporation} database attached to \textsc{Geant4}, which guarantees data consistency at each step with another step, and with the normal, straightforward simulation.
This data consistency is an advantage over MGGPOD \cite{Weidenspointner:2005}, which is a combination of different simulation codes, and our previous work \cite{Mizuno:2010} which combined MGGPOD and Geant4.
MGGPOD actually handled a single isomeric level of each isotope in the isotope generation phase, which may be based on an assumption that a ground state and at most one isomeric level have significant contributions to the background spectra from an isotope species.
This may not be true in reality, and an accurate treatment of isomers should be the key to the model accuracy.

Another important point newly introduced in this work is the treatment of the entire history of the particle irradiation.
Since the particle flux is highly variable along the orbit (see Figure~\ref{fig:particle_monitor}), the delayed nature of the background resulting from radioactivation requires treatment of arbitrary time profile of irradiation at Step 2 in our framework.
The key difference of this work from previous codes such as MGGPOD \cite{Weidenspointner:2005} and MEGAlib \cite{Zoglauer:2006} is that we can now easily handle arbitrary time profiles of irradiation and observation which would be much more difficult to achieve with them.
We described in \S\ref{sec:method} the formulation of the arbitrary time profiles of the irradiation and the measurement using the analytical solutions, Eqs.~(\ref{eq:solution1}), (\ref{eq:solution2}), of the Bateman equation.
We use the Bayesian block algorithm \cite{Scargle:2013a,Scargle:2013b} for data compression of the time profile data obtained with the particle monitor.
This algorithm reduces the count rate data into a smaller number of time sections within which count rates are kept constant, giving a statistically correct way to handle a long time profile data in our semi-analytical approach at Step 2.
The degree of compression is adjustable by a hyperparameter of the algorithm.

\begin{figure*}[t]
\begin{center}
\includegraphics[width=18cm]{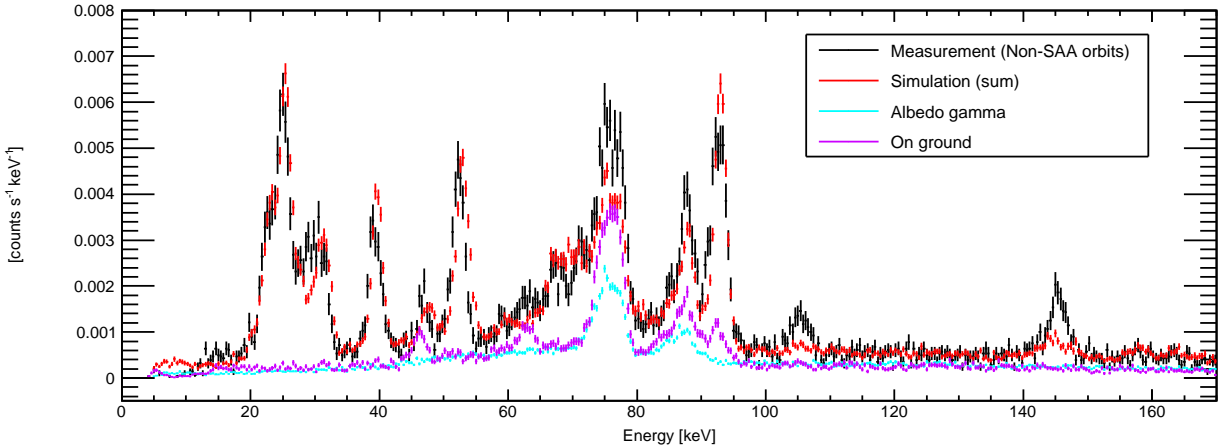}
\caption{Comparison of the in-orbit measurement in the non-SAA orbits, the simulation of the in-orbit background, and the ground measurement. Note that the simulation does not include on-ground measurement. The simulation of the albedo gamma rays is also plotted to show which lines are dominated by the bismuth fluorescence.}
\label{fig:ground_measurement}
\end{center}
\end{figure*}

As we already mentioned, disagreement remains at some line energies.
Some of those lines can be attributed to radioactive isotopes that were naturally contaminated in materials in the detector system before the  launch of the satellite.
Figure~\ref{fig:ground_measurement} shows a background spectrum measured on the ground (before the launch), making comparison with the measurement and the simulation of the in-orbit background in the non-SAA orbits.
This ground measurement includes the natural contamination as well as external background radiation from outside of the detector system in the measurement environment.
This external component only contributes to the fluorescence of bismuth as the CXB and albedo gamma rays generate background only via fluorescence.
Thus, the rest of the lines seen in the ground measurement at 46 keV, 63 keV, and 92 keV originate from the contamination of  isotopes, explaining the residuals between the simulation and the data at these energies.
Furthermore, the residuals at the fluorescence energies (75--77 keV and 87 keV) can also be attributed to the internal origin since radiations by the contaminated isotopes should generate the bismuth fluorescence.

Still, the simulations have significant differences from the data at a few line energies.
Lines in the simulations at 106 keV and 145 keV are obviously too weak compared to the measured spectra both in the SAA and non-SAA orbits.
These two lines are certainly produced in the simulation (see the logarithmic scale plots in Figures~\ref{fig:background_spectra_saa} and \ref{fig:background_spectra_nonsaa}), but their intensities are too low by a factor of 4.
The simulations identify the lines to $^{129m}\mathrm{Te}$ (105.5 keV) and $^{125m}\mathrm{Te}$ (144.8 keV), both of which are due to isomeric transitions.
Both those isomers have relatively long half-lives comparable to a month, which is longer than the measurement time windows, and their lines do not show time variability between the measurements.
This means that they are produced by the primary interactions, not via short-lived isotopes.
Thus, the cascade model in the inelastic interactions of the primary protons fails to produce a sufficient amount of the isomers.

Another discrepancy is seen at 157 keV in the SAA orbit spectrum, which is identified as $^{122m}\mathrm{In}$ (156.6 keV).
This line is, by contrast to the previous almost missing lines at 106 keV and 145 keV, 2--3 times stronger in the simulation than in the measurement.
A major part of the reasons can be explained by the instrumental response.
This energy can exceed the upper limit of the analog-to-digital converter range, depending on a value of common mode noise \cite{Hagino:2017}.
Thus, a trigger efficiency at this energy is estimated to be $\sim$60\%.
Taking account of this efficiency suppression, the simulation-to-measurement ratio gets even better.
The remaining difference is less than a factor of 2 and could be attributed to the uncertainties of the physics model.
We should also note that this isomer has a short half-life of 21 minutes, and therefore its intensity could differ if the irradiation time profile was distorted. Uncertainties of the measured time profile are discussed below.

Evaluation of the atomic line complex at 25 keV is not trivial.
This comes from atomic transitions of different isotopes generated via radioactivation of cadmium nuclei.
Although the intensity of this line feature agrees very well with the measurement in the non-SAA orbits, the simulation overestimates its intensity in the SAA orbits.
This means that the decay constant averaged over the relevant isotopes is different between the simulation and reality, which implies that some short-lived isotopes that contribute to the atomic complex might be overproduced in the simulation.



In addition to uncertainties of the hadron physics model and its relevant databases, the radiation environment as an input to the simulation must have uncertainties.
Although we used the particle monitor data for relative amplitude of the proton irradiation, the absolute normalization is determined by the AP-8 model (see \S\ref{subsec:enviroment}), which can have an uncertainty of $\sim$50\% \cite{Armstrong:2000}.
It may be worthwhile to consider another proton flux model based on recent measurements such as \textit{PAMELA} \cite{Adriani:2015, Bruno:2017}.
In addition, we used a time-averaged spectrum and an isotropic directional distribution for the trapped protons, but this could be too simple since the radiation belt has complicated structure.
Furthermore, it is possible that the particle monitor could not fully trace the time variability of the input radiation at the outside of the entire detector system, since the monitor APD was placed inside the thick BGO shields with a translation offset and a perpendicular direction to the main imager.


Finally, we suggest a few points to improve the model accuracy for future hard X-ray missions.
Although we already have a simulation framework that will be applicable to various types of future missions, the hadron physics model and its relevant databases should be verified by measurement (e.g., ground-based tests at proton-beam facilities).
It is extremely useful to conduct a series of high-precision measurements at different time-scales (to cover a broad lifetime range) after a proton beam irradiation of materials used in instruments \cite{Murakami:2003} for calibrating the model and databases.
The input radiation to the simulation should also depend more extensively on the measurement.
It will be useful to obtain more information (e.g., absolute flux, spectrum, and anisotropy) on the primary particles based on in-orbit measurement.

\section{Conclusions}
\label{sec:conclusions}

We have developed a new, effective, general-purpose framework of full Monte Carlo simulation of radioactivation background in hard X-ray detectors induced by energetic particles such as cosmic-ray protons and/or geomagnetically trapped protons.
In order to efficiently treat the delayed nature of the radioactivation background, we separate the entire processes into the phase of generation of radioactive isotopes by primary irradiations and the phase of decays of those generated isotopes and their descendants at measurement time, inserting newly developed semi-analytical conversions of production probabilities of those activated isotopes into the decay rates at the time of interest. 
We use the publicly available \textsc{Geant4} toolkit library and its attached databases in all steps of our simulations and calculations, which as a result benefits from database consistency through all the steps and from full compatibilities with other classes of background simulations (e.g., cosmic X-ray background, albedo gamma rays and albedo neutrons).
We applied this framework to the case of a CdTe semiconductor detector, and verified that our method produced results consistent with normal, straightforward Monte Carlo simulations, but achieved 100-fold reduction of computation time.

We then considered in-orbit background spectra obtained with the CdTe imaging sensor of the HXI flown on the \textit{Hitomi} X-ray astronomy satellite for evaluating experimental performance of the simulation framework.
As this satellite was put into LEO, the radioactivation is induced dominantly by geomagnetically trapped protons associated with the SAA, where the irradiation rate had been monitored by a particle counter placed in close proximity to the main focal plane detector.
Assuming the input spectrum and absolute flux of the trapped protons calculated by the AP-8 model via \textit{SPENVIS}, 
we calculated the background spectra using our simulation framework which considered the entire history of the proton irradiation, and compared its results to the measured data.
The agreement between the simulated and measured spectra is excellent, demonstrating that the instrumental background of the CdTe detector of the HXI consists of radioactivation of CdTe itself and surrounding BGO active shields, photon leakages of cosmic X-ray background and Earth's albedo gamma rays through openings of the detector shields and baffles, and emissions from naturally contaminated radioactive isotopes inside the instrument components.
Still remaining discrepancies can be attributed to imperfect hadron physics models and databases of the simulation and to uncertainties of the input radiation environments.
Efforts to improve these factors by exploiting more precise measurements will be necessary for future hard X-ray missions.

\section*{Acknowledgement}

We acknowledge all the JAXA members who have contributed to the ASTRO-H (Hitomi) project. Stanford and SLAC members acknowledge support via DoE contract to SLAC National Accelerator Laboratory DE-AC3-76SF00515, as well as the NASA grant NNX15AM19G to support the Science Enhancement Program for Astro-H (Hitomi) at Stanford.  


\bibliography{hxi_simulation_references}

\end{document}